\documentclass[twocolumn]{aastex62}

\usepackage{multirow}
\usepackage{booktabs}
\usepackage{graphicx}
\usepackage{amsmath}
\usepackage{longtable}
\usepackage{rotating}
\usepackage{enumerate}
\usepackage{color}

\newcommand{\cs}{c_{\rm s}}

\newcommand{\vth}{v_{\rm th}}
\newcommand{\vk}{v_{\rm K}}
\newcommand{\vlos}{v_{\rm LOS}}
\newcommand{\tmid}{T_{\rm midplane}}
\newcommand{\msun}{M_\odot}
\newcommand{\mj}{M_{\rm J}}
\newcommand{\rp}{r_{\rm p}}
\newcommand{\mplanet}{M_{\rm p}}

\begin{document}
\title{Observational Signatures of Planets in Protoplanetary Disks: Planet-Induced Line Broadening in Gaps}

\author{Ruobing Dong}
\affil{Department of Physics \& Astronomy, University of Victoria, Victoria, BC, V8P 1A1, Canada}

\author{Sheng-Yuan Liu}
\affil{Institute of Astronomy and Astrophysics, Academia Sinica, 11F of ASMAB, AS/NTU No.1, Sec. 4, Roosevelt Rd, Taipei 10617, R.O.C.}

\author{Jeffrey Fung}
\altaffiliation{NASA Sagan Fellow}
\affil{Department of Astronomy, University of California at Berkeley, Campbell Hall, Berkeley, CA 94720-3411, USA}

\clearpage

\begin{abstract}

Protoplanets can produce structures in protoplanetary disks via gravitational disk-planet interactions. Once detected, such structures serve as signposts of planet formation. Here we investigate the kinematic signatures in disks produced by multi-Jupiter mass ($\mj$) planets using 3D hydrodynamics and radiative transfer simulations. Such a planet opens a deep gap, and drives transonic vertical motions inside. Such motions include both a bulk motion of the entire half-disk column, and turbulence on scales comparable to and smaller than the scale height. They significantly broaden molecular lines from the gap, producing double-peaked line profiles at certain locations, and a kinematic velocity dispersion comparable to thermal after azimuthal averaging. The same planet does not drive fast vertical motions outside the gap, except at the inner spiral arms and the disk surface.
Searching for line broadening induced by multi-$\mj$ planets inside gaps requires an angular resolution comparable to the gap width, an assessment of the gap gas temperature to within a factor of 2, and a high sensitivity needed to detect line emission from the gap.
\end{abstract}

\keywords{protoplanetary disks --- stars: variables: T Tauri, Herbig Ae/Be --- planets and satellites: formation --- planet-disk interactions --- planets and satellites: detection}


\section{Introduction}\label{sec:intro}

Detecting protoplanets forming in protoplanetary disks provides fundamental constraints on key parameters governing the planet formation processes, such as location, timescale, and local environment. 
Apart from directly detecting the photosphere, protoplanets can be detected through emissions from their circumplanetary disks \citep[CPD; e.g.,][]{zhu15cpd, eisner15, szulagyi17}, as well as planet-induced kinematics \citep[e.g.,][]{perez15cpd, perez18} and chemical signatures in the circumstellar disk \citep{cleeves15}.
Massive efforts have been carried out to look for protoplanets via direct imaging; yet, most yielded null results \citep[e.g.,][]{testi15, canovas17, maire17, uyama17twhya}. 
In a few cases, candidate point sources have been detected \citep[e.g.,][]{quanz13planet, currie14, sallum15, reggiani18, guidi18, keppler18}, while in most of them the nature of the detection is under debate (e.g., bona fide planets {\it vs} disk structures; \citealt{thalmann16, follette17, hord17, mendigutia18}).

In this paper, we investigate a large scale kinematic signature that a forming Jovian planet produces in a disk --- meridional gas motions in planet-opened gaps \citep{morbidelli14, szulagyi14, fung16}. 
The mechanism of gap opening can be described as the balance between the planetary torque that repels material from its co-orbital region and the viscous torque that attempts to fill it back in \citep{lin93}.
Previous work have shown that under typical conditions, a 1 Jupiter mass ($\mj$) planet can open a deep gap a few orders of magnitude depleted in gas \citep[e.g.,][]{fung14,fung16}.
Under hydrostatic equilibrium, the gas density distribution $\rho(r, z)$ in the vertical direction $z$ at a stellocentric radius $r$ is
\begin{equation}
\label{eq:rhoz}
\rho(r, z)=\rho_0(r)e^{-\frac{z^2}{2h(r)^2}},
\end{equation}
where $\rho_0(r)$ is the midplane density at $r$
, and $h(r)$ is the disk scale height at $r$.
Close to the planet's orbit $\rp$, $h$ is nearly constant. 
The planet's torque is strongest near the disk midplane ($z=0$), and diminishes with increasing $|z|$. The viscous torque is on the other hand more evenly distributed. Even though these two torques are {\it overall} balanced, they are unable to cancel each other locally. This leads to a planetary torque dominated midplane where gas flows away from the planet, and a viscous torque dominated upper atmosphere that flows toward the planet. The two flows connect via meridional circulations, which can reach velocities comparable to local sound speed $\cs$ in gaps opened by super-Jupiter planets \citep{fung16}. These gas motions may significantly broaden molecular emission lines. This is the key signature that we will examine. 

The main motion of gas in a disk is Keplerian rotation with a velocity $\vk$. Significant deviations from Keplerian rotation may indicate the presence of substantial density structures \citep[e.g., a warp, ][]{rosenfeld14, casassus15gas, facchini18, mayama18}. The thermal motion of gas molecules has a typical velocity of $\vth\sim\cs\sim(h/r)\vk$, usually several percent of $\vk$. Instabilities, such as the magnetorotational instability (MRI), may induce sub-sonic turbulent motions \citep[e.g.,][]{simon11}.

Planet-induced kinematic signatures have been investigated before. In a pioneering work, \citet{perez15} assessed the detectability of the rotation of CPDs. More recently, \citet{pinte18} showed that a planet may introduce local kinks in the isovelocity maps of the circumstellar disk, \citet{teague18} quantified the radial modulation in the circumstellar disk rotation due to gap structures, and \citet{huang18vortex} studied the detectability of the anticyclonic gas motion inside planet-triggered vortices. \citet{perez18} investigated gas kinematics caused by pressure gradients at gaps, spiral wakes and vortices. While these works mainly focused on planet-induced signatures in intensity weighted velocity (moment-1) maps\footnote{See \citet{teague18_centroids} for a discussion on inferring line centroids in real observations}, we study how meridional motions are manifested in velocity dispersion (moment-2) maps.

Today, Keplerian rotation can be probed precisely \citep[e.g., used to infer the mass of the central star(s);][]{czekala15, czekala16, czekala17}. Searching for sub-sonic turbulent motions such as MRI turbulence \citep[e.g.,][]{bai15, simon15, simon17} is an ongoing effort. Pioneering efforts have been carried out using various molecule tracers, such as \citet[CO, TW Hya and HD 163296]{hughes11}, \citet[CS, DM Tau]{guilloteau12}, 
\citet[CO,$^{13}$CO, C$^{18}$O, HD, 163296]{flaherty15}, 
\citet[CO, CN, and CS, TW Hya]{teague16}, \citet[DCO$^+$, C$^{18}$O and CO, HD 163296]{flaherty17}, 
\citet[CO, TW Hya]{flaherty18},
and \citet[CS, TW Hya]{teague18twhya}. 
Such efforts now can typically constrain turbulent velocities $v_{\rm turb}$ to $<10\%\cs$.

Our goal --- to study the observational signatures of meridional motions in planet-opened gaps --- is similar in nature to those in looking for turbulence generated by instabilities. We combine 3D hydrodynamics (\S\ref{sec:hydro}) and radiative transfer simulations (\S\ref{sec:rt}) to produce synthetic line emission observations. We illustrate the key observational signature inside the gap (\S\ref{sec:basic}), while commenting on the variations in the key signature under various circumstances (\S\ref{sec:variations}) as well as gas motions outside the gap (\S\ref{sec:outsidegap}). A generic strategy in searching for such signatures in real observations is outlined (\S\ref{sec:applications}), and advantages of those signatures as signposts of planets are highlighted (\S\ref{sec:advantages}).


\section{Simulations}\label{sec:setup}

We adopt a general scheme to produce synthetic line observations similar to \citet{perez18}. First, we carry out 3D global hydrodynamics simulations using the \texttt{PEnGUIn} code \citep{fung15thesis} to calculate the density and velocity structures of disks perturbed by a giant planet. Then, we combine dust radiative transfer simulations using \texttt{HOCHUNK3D} \citep{whitney13} and molecular line radiative transfer simulations using the package "Simulation Package for Astronomical Radiative Xfer" (\texttt{SPARX}\footnote{ https://sparx.tiara.sinica.edu.tw/}) to produce channel maps.

\subsection{Hydrodynamics Simulations}\label{sec:hydro}

The setup we use for our hydrodynamics simulations of gas-opening largely borrows from \citet{fung16}. We perform simulations in spherical coordinates, and the grid has dimensions of 320 (radial) by 810 (azimuthal) by 50 (polar) cells. Our domain extends from $0.3\rp$ to $2\rp$ radially in logarithmic spacing, covers the full $2\pi$ in azimuth, and includes 0.25 radians from the midplane to the top of the grid in polar angles. We simulate only one half of the disk, taking advantage of its symmetry around the midplane. This grid setup is nearly identical to the one used by \citet{fung16}, except both the radial and polar domain have been extended to improve our ability to isolate the gap and resolve the disk surface. The resolution we choose have been shown to give good convergence on the gap profile \citep{fung14,fung16}. Since the gap profile is determined by the balance of planetary torque and disk diffusion, this implies our resolution is also sufficient to constrain the amount of turbulent diffusion in the system.

For boundary conditions, in the radial direction we have fixed boundaries with wave killing zones; in the azimuthal direction it is periodic; and in the polar direction it is reflective, both at the midplane to enforce symmetry, and at the top to prevent unrealistic mass influx or outflux. Also, we use an updated version of \texttt{PEnGUIn} \citep{fung18} that includes the fast orbital advection algorithm \citep{masset00}, which helps speed up the code and reduce numerical diffusion.

We carry out two simulations each with one planet on a fixed circular orbit at $\rp=30$ AU for 1000 planetary orbits. The planets each has a mass of $\mplanet=10^{-3}M_\star$ (Model Q1) and $4\times10^{-3}M_\star$ (Model Q4); i.e., 1~$\mj$ and 4~$\mj$ if $M_\star=1\msun$.
The disk has an initial surface density profile of:
\begin{equation}
\Sigma = \Sigma_0 \left(\frac{r}{\rp}\right)^{-\frac{1}{2}} \, ,
\end{equation}
and a time-invariant vertically
isothermal temperature profile that gives a sound speed of:
\begin{equation}
\label{eq:hydro_cs}
\cs = 0.05 \vk\left(\frac{r}{\rp}\right)^{-\frac{1}{2}} \, .
\end{equation}
This choice of using a vertically isothermal disk is inspired by the fact that at tens of AU the disk is expected to be vertically isothermal for bulk of its mass. A more realistic equation of state (EOS) may introduce an order-of-unit correction to our models, but unlikely to alter our results qualitatively.

Our simulations do not account for disk self-gravity, thus we are free to tune $\Sigma_0$ to set the total disk mass $M_{\rm disk}=0.001\msun$ in both models. We choose this $M_{\rm disk}$ so that the disk is globally optically thin to $^{13}$C$^{18}$O emission (\S\ref{sec:basic}). 
We include a small viscosity using the \citet{shakura73} prescription, $\nu = \alpha h \cs$, where $\nu$ is the kinematic viscosity, and $\alpha$ is set to $10^{-3}$. This amount of viscosity ensures that the disk reaches steady state within 1000 planetary orbits. Note that physically this viscosity corresponds to turbulent diffusion, which would manifest itself in line broadening. This ``viscous turbulence'' is not included in our line emission simulations, as we focus on planet-induced effects only (non-planet-induced turbulence in disks is likely weak, see \S\ref{sec:intro}). In our plots, the disk rotates counterclockwise.

Naturally, simulation results greatly resemble those by \citet{fung16} (Figure~\ref{fig:basic}($a$)($b$)($c$)). In Appendix \ref{app:flow}, Figure \ref{fig:v_flow_2p} can be compared to Figure 4 of \citet{fung16}, and Figure \ref{fig:meridionalkinematicenergy} illustrates the distribution of kinetic energy due to planet-induced vertical motions in the disk. While we are originally motivated by the meridional circulation in and around the gaps, we find that the gap in model Q4 is turbulent, and this turbulent motion can contribute to line broadening even more greatly than the circulation (see \S\ref{sec:discussions} for more discussions).

\begin{figure*}
\begin{center}
\text{\large Surface Density and Temperature Structures in Our Models} \par\smallskip
\includegraphics[trim=0 0 0 0, clip,width=\textwidth,angle=0]{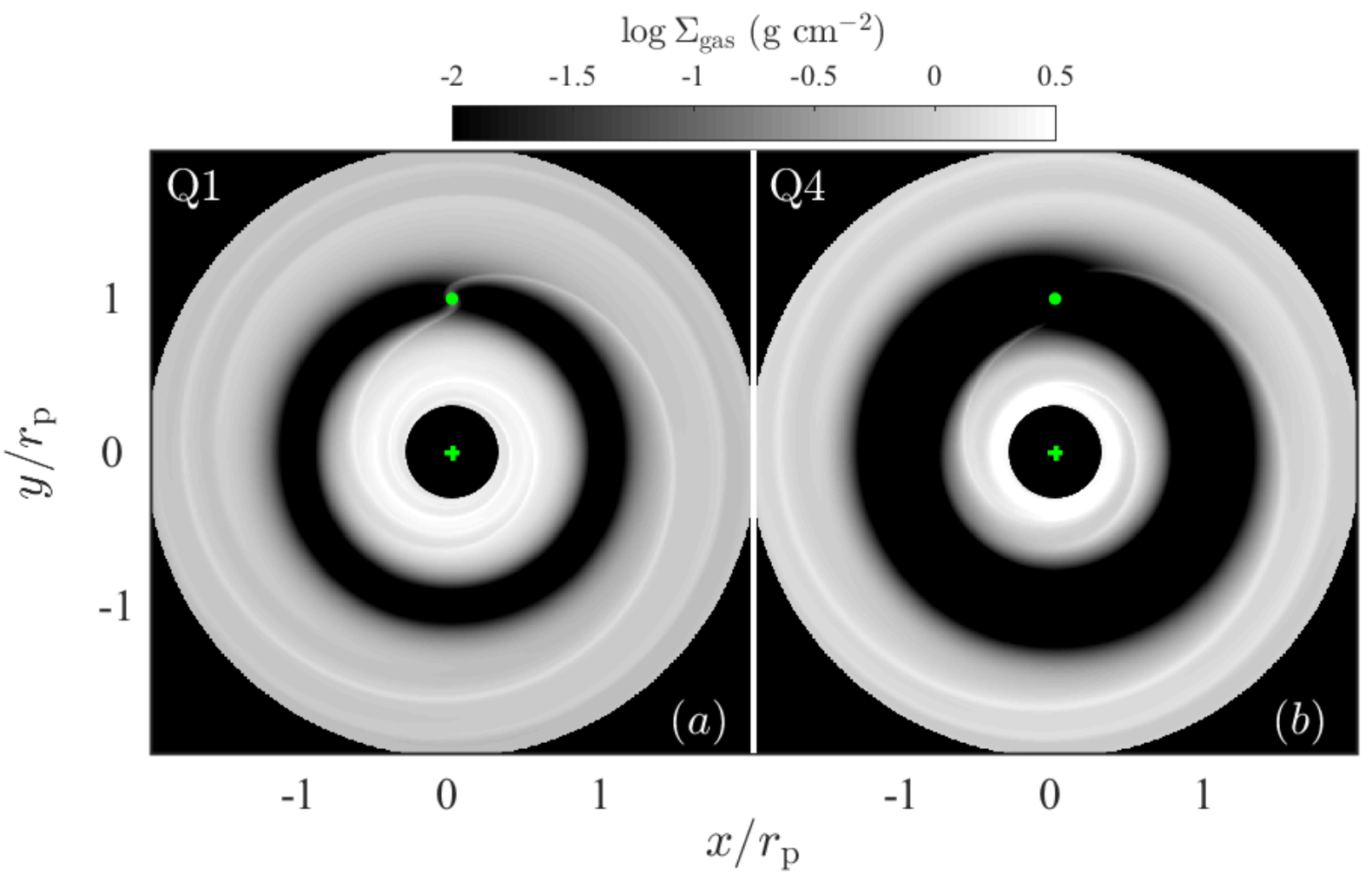}
\includegraphics[trim=0 0 0 0, clip,width=0.32\textwidth,angle=0]{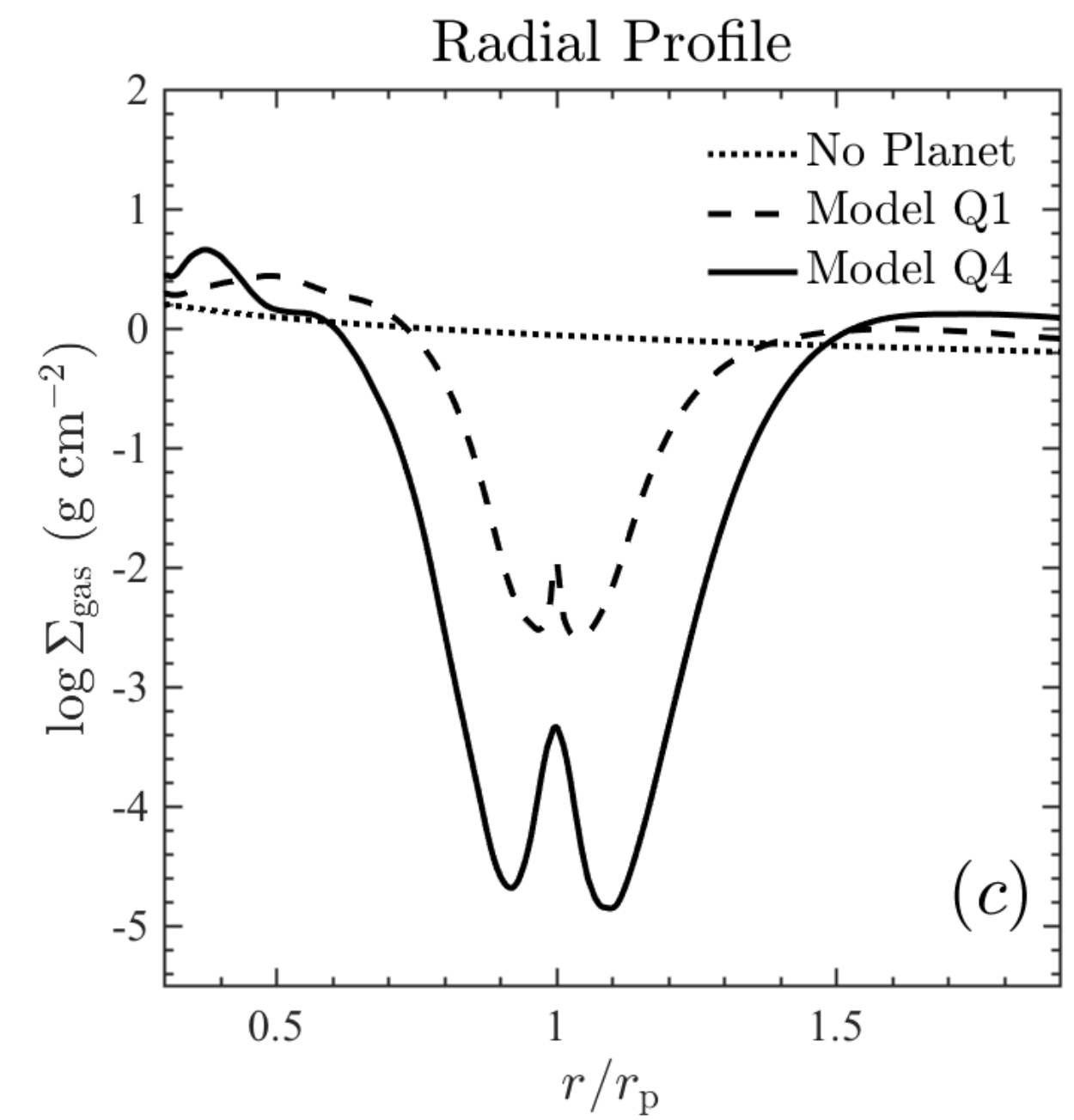}
\includegraphics[trim=0 0 0 0, clip,width=0.32\textwidth,angle=0]{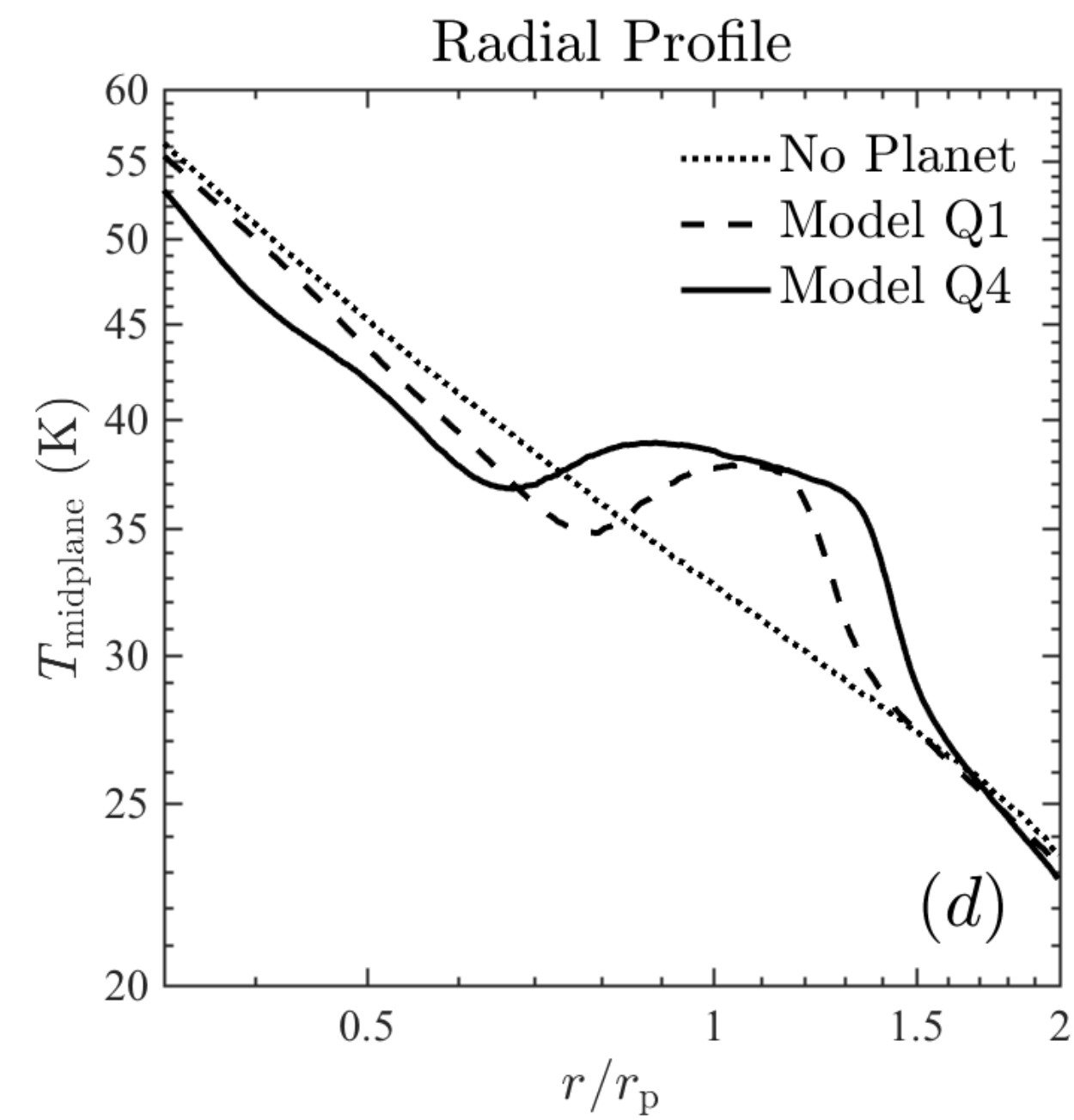}
\includegraphics[trim=0 0 0 0, clip,width=0.32\textwidth,angle=0]{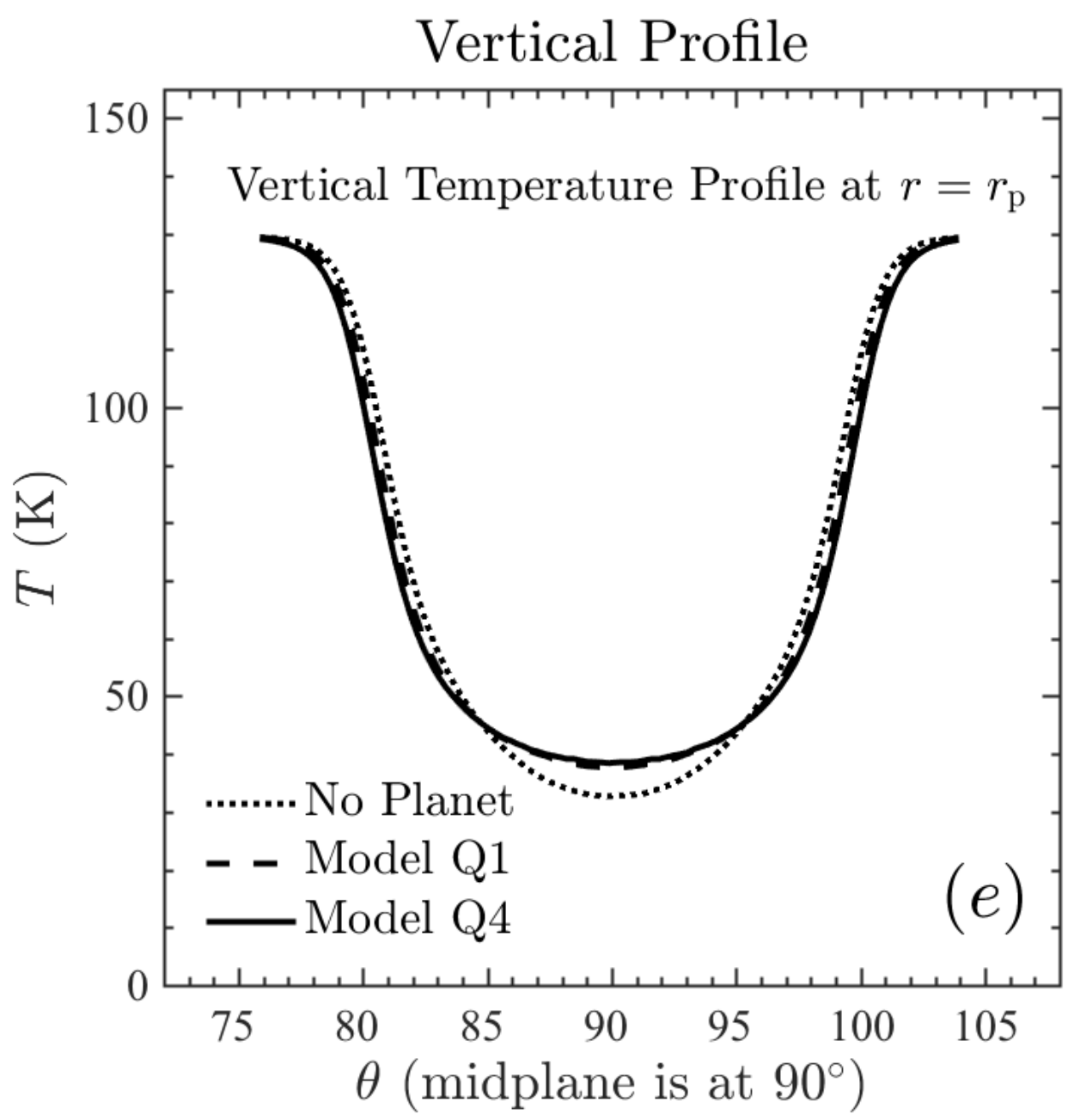}
\end{center}
\figcaption{{\it Top}: Surface density maps of the two models at 1000 planetary orbits (($a$) and ($b$)); the green plus and dot mark the locations of the star and the planet, respectively. {\it Bottom:} Azimuthally averaged radial profiles of the surface density ($c$) and midplane temperature ($d$), and azimuthally averaged vertical temperature profiles at $\rp$ (($e$); $\theta$ is the polar angle). See \S\ref{sec:setup} for discussions.
\label{fig:basic}}
\end{figure*}

\subsection{Radiative Transfer Simulations}\label{sec:rt}

We feed the exact 3D hydro density structures into \texttt{HOCHUNK3D} to calculate the disk temperature, following the procedures described in \citet{dong16armviewing}. The central star, which is the only heating source in our models, is assumed to be a 1 $\msun$ protostar with a temperature of 4350 K and a radius of 2.3 $R_\odot$. A population of interstellar medium (ISM) dust \citep[sizes up to $\sim1\micron$]{kim94} is employed to process starlight; they are assumed to be well-mixed with the gas with their density set to 0.1\% of the gas density. Both dust scattering (anisotropic; scattering phase function approximated using the Henyey--Greenstein function) and absorption/emission are included (dust properties can be found in Figure 2 in \citealt{dong12cavity}).

Figure~\ref{fig:basic}($d$) shows azimuthally averaged midplane temperature ($\tmid$) for both models, as well as in a control model with no planet (i.e., a full disk). The gap region is on average about 5 K, or 15\%, hotter than in the no planet case. Panel ($e$) shows azimuthally averaged vertical temperature profiles at $\rp$. The midplane is the coldest, and $T$ rises with $z$ before plateauing at the surface. A nearly isothermal midplane layer with $T-\tmid<10$ K encompasses $\sim95$\% of the mass.

We note that temperature, and therefore the vertical hydrostatic structure, in our models are not self-consistent. Hydrodynamics simulations assume a {\it parametrized} scale height profile (Equation \ref{eq:hydro_cs}), which under the hydrostatic equilibrium condition corresponds to a midplane temperature not fully consistent with the one obtained from radiative transfer (RT) simulations (the former is $\sim$30\% lower at $\rp$ in the no planet case; note in self-consistent disk models the difference may be higher, \citealt{isella18}). 
Achieving self-consistency requires either coupled hydro-RT simulations, or iterations, which are beyond our scope. This self-inconsistency however does not affect the key observational signature to be defined in the next section (see more discussions in \S\ref{sec:variations}).

Synthetic molecular line channel maps and moment maps are produced with \texttt{SPARX}, which calculates the spectral line radiative transfer of specific transitions for molecules of interests. Molecular gas densities and velocities extracted from the hydrodynamics simulations and disk temperature derived from the \texttt{HOCHUNK3D} simulations are provided to \texttt{SPARX}. For demonstration purposes, we consider the rotational transitions in $J=3-2$ of CO and its isotopologues C$^{18}$O, and $^{13}$C$^{18}$O.
Fractional abundances of CO, C$^{18}$O, and $^{13}$C$^{18}$O relative to H$_2$ are set at 10$^{-4}$, 2 $\times$ 10$^{-7}$, and 2 $\times$ 10$^{-9}$, respectively. These transitions are mainly for covering different levels of opacity, and the listed abundances roughly follow the measurements in ISM dense clouds \citep[e.g.,][]{vandishoeck98, wilson94}. 

While SPARX is capable of evaluating iteratively the non-local-thermodynamical-equilibrium (non-LTE) molecular level populations and the corresponding radiation intensity, the high gas density throughout the disk domain under consideration is sufficient to thermalize the $J=3-2$ transitions of CO and its isotopolouges, whose critical densities are around $10^{4} - 10^{5}$ cm$^{-3}$. Thus, we applied the LTE level population following the Boltzmann distribution and performed standard ray-tracing for generating the synthetic images. 

A pixel dimension of 300$\times$300 and a velocity resolution of 0.03 km~s$^{-1}$ are used for ensuring the molecular emission features are both spatially and spectrally resolved. Dust continuum emission is included in our line radiative transfer calculations; it is insignificant due to the low $\sim$mm-wavelength opacity of the assumed ISM dust (on the order of 0.1 cm$^2$/g). Millimeter opacity in real disks is expected to be 1--2 orders of magnitude larger; our choice helps isolate and reveal the planet-induced signatures in the cleanest way. Simulated channel maps are not corrupted to incorporate realistic noises and angular resolutions. We defer direct model-observation comparisons to future work.

Finally, we note that moment-2 of line emission from optically thin gas at temperature $T$ with no non-thermal motions equals to the isothermal sound speed $\cs$,
\begin{equation}
\label{eq:thermal}
{\mbox{Moment-2}}=\cs=\sqrt{\frac{kT}{m_\mu}}=94\sqrt{\frac{T}{\rm 30K}\frac{m_{\rm CO}}{m_\mu}}\ \rm m/s
\end{equation}
where $m_\mu$ is the molecular weight.


\section{The Key Observational Signature of Planet-Induced Velocity Dispersion}\label{sec:basic}

In this section, we define and study the key observational signature of planet-induced turbulence inside the gap in Model Q4. We employ a tracer, $^{13}$C$^{18}$O ($J=3-2$), to which the disk is globally optically thin (optical depth $\tau\sim0.1$ at $v_{\rm LOS}=0$ outside the gap and $\tau$ peaks at $\sim10^{-7}$ inside the gap). We adopt a face-on geometry, so that the vertical direction in the disk corresponds to the line of sight (LOS), and vertical velocity dispersion corresponds to moment-2. The dependence of the key signature on parameters will be discussed in \S\ref{sec:variations}.

We emphasize that experiments in this section (and most of the paper) are designed to demonstrate the basic principle, rather than to mimic real life situations. In particular, we choose $^{13}$C$^{18}$O, a species almost impossible to detect inside a deep gap at the moment, 
because it gives the most simple picture of the planetary signature.
This simplicity comes from the fact that the thermal broadening would be a smooth and almost monotonic function with radius if, and only if, the emission originates from the midplane everywhere. This makes the planet-induced broadening stands out more clearly.
As we will see (\S\ref{sec:variations}; Figure~\ref{fig:moment2_co}), planet-induced line broadening inside gaps is clearly present (and retrievable; \S\ref{sec:applications}) in emission from tenuous gas tracers such as CO, as long as the gap is optically thin to the emission, which is often easy to satisfy due to the large gap depletion.

\begin{figure}
\begin{center}
\text{\large Model Q4, $\rp=30$ AU, $^{13}$C$^{18}$O, Face On} \text{\large (reference: $\cs(40\rm K)=0.10$ km/s)} \par
\includegraphics[trim=10 0 0 0, clip,width=0.5\textwidth,angle=0]{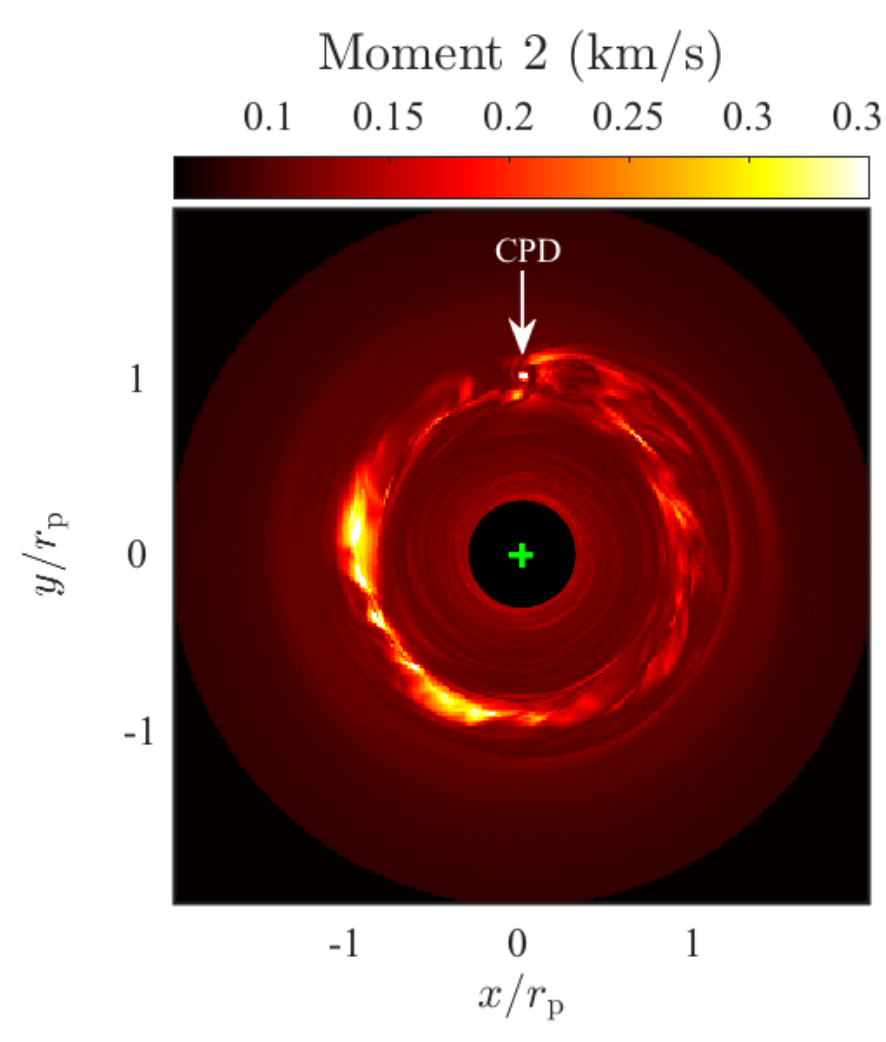}
\end{center}
\figcaption{Moment-2 in Model Q4 for $^{13}$C$^{18}$O $J=3-2$ transition at face on. The location of the circumplanetary disk (CPD) is labeled. A $4\times10^{-3}M_\star$ planet induces a prominent ring of high velocity dispersion inside the gap. This is the key planet-induced signature studied in this paper. See \S\ref{sec:basic} for details.
\label{fig:moment2_4mj}}
\end{figure}

Figure~\ref{fig:moment2_4mj} shows the moment-2 map. A ring of high velocity dispersion around the planet's orbit is prominent. This is the key signature of planet-induced turbulence.

\begin{figure*}
\begin{center}
\text{\large Model Q4, Moment-2, $^{13}$C$^{18}$O, Face On (reference: $\cs(40\rm K)=0.10$ km/s)} \par\smallskip
\includegraphics[trim=0 0 0 0, clip,width=\textwidth,angle=0]{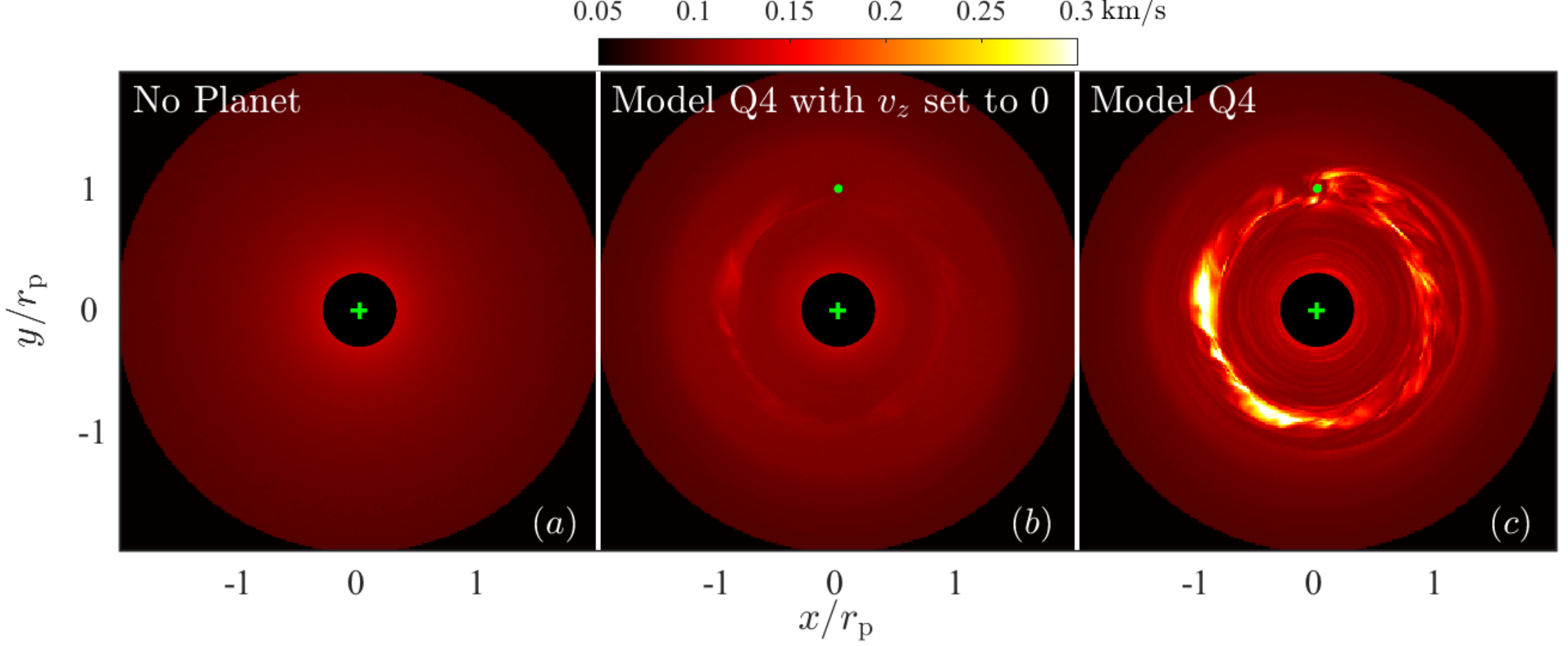}
\includegraphics[trim=0 0 0 0, clip,width=0.5\textwidth,angle=0]{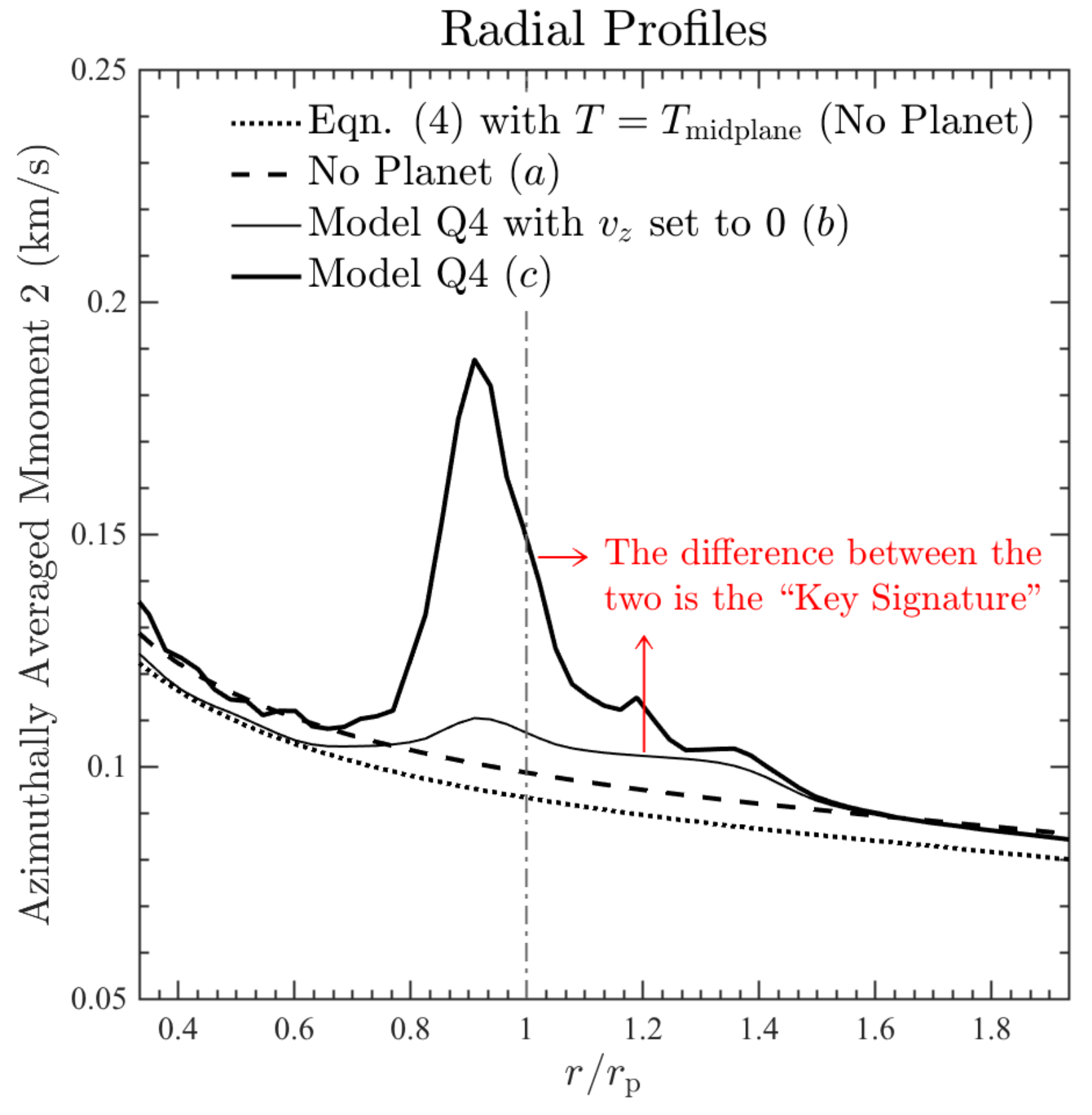}
\end{center}
\figcaption{Moment-2 in $^{13}$C$^{18}$O $J=3-2$ transition at face on in the no planet case ($a$) and in Model Q4 ($c$). Panel ($b$) shows the map for Model Q4 calculated with the vertical velocity $v_z$ manually set to 0. The contrast between ($b$) and ($c$) highlights the effects of planet-induced vertical motions. The bottom panel shows azimuthally averaged radial profiles. The large velocity dispersion inside the gap (the ``bump'' at $\sim0.9\rp$ on the Model Q4 profile) is the key planet-induced kinematic signature. See \S\ref{sec:basic} for details.
\label{fig:moment2_basic}}
\end{figure*}

Figure~\ref{fig:moment2_basic} compares moment-2 in the no planet case ($a$) with that in Model Q4 ($c$). In the former, the velocity dispersion is entirely thermal --- Eqn.~\ref{eq:thermal} with $T=\tmid$ (dotted line in the bottom panel) closely traces the measured moment-2 (dashed line). The small difference can be attributed to the higher temperatures of the tenuous gas at high latitudes. In the latter, the velocity dispersion outside the gap (roughly $0.8-1.5\rp$) is consistent with thermal broadening too, suggesting no substantial planet-induced motions there. 

Inside the gap, however, there is a ring of high velocity dispersion. After azimuthal averaging, at its peak at $\sim$0.9$\rp$ the total line broadening is about twice of the thermal broadening, indicating a ``mean'' non-thermal velocity of $\sim$1.7$\cs$ (total $\approx\sqrt{\rm thermal^2+turbulent^2}$ for Gaussian-like turbulence). To highlight the ``planet-induced'' nature of this additional broadening, panel ($b$) shows the moment-2 map simulated with the vertical velocity $v_z$ manually set to 0, so that only thermal broadening contributes. In this case, moment-2 inside the gap is only $\sim$10\% higher than that in the no planet case due to the slightly higher temperature in the former (Figure~\ref{fig:basic}). 
Inside the planet's Hill radius (marked in Figure~\ref{fig:moment2_4mj}; resolved by $\sim$15 cells in hydro simulations) we observe a velocity dispersion exceeding 30\%$\vk$, almost entirely non-thermal (note that the heating from the CPD accretion is not taken into account).

\begin{figure*}
\begin{center}
\text{\large Model Q4, $^{13}$C$^{18}$O, Face On (reference: $\cs(40\rm K)=0.10$ km/s)}
\par\smallskip
\includegraphics[trim=0 0 0 0, clip,width=0.32\textwidth,angle=0]{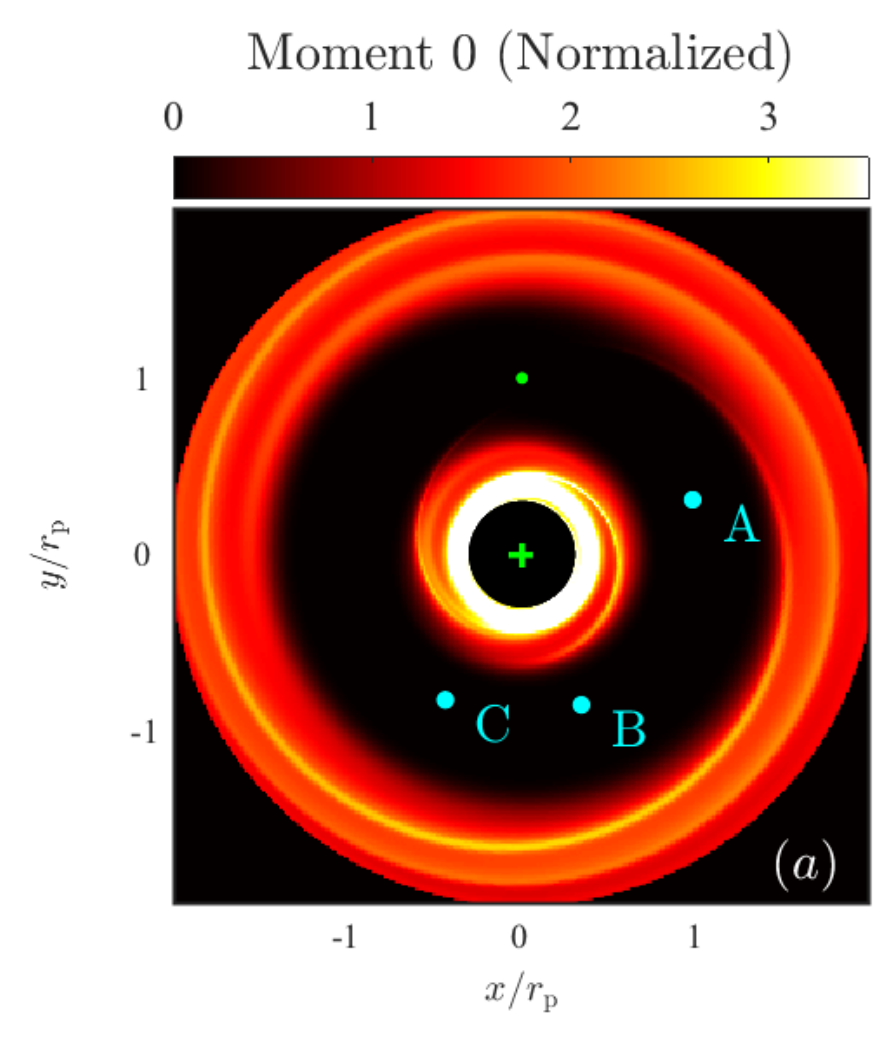}
\includegraphics[trim=0 0 0 0, clip,width=0.32\textwidth,angle=0]{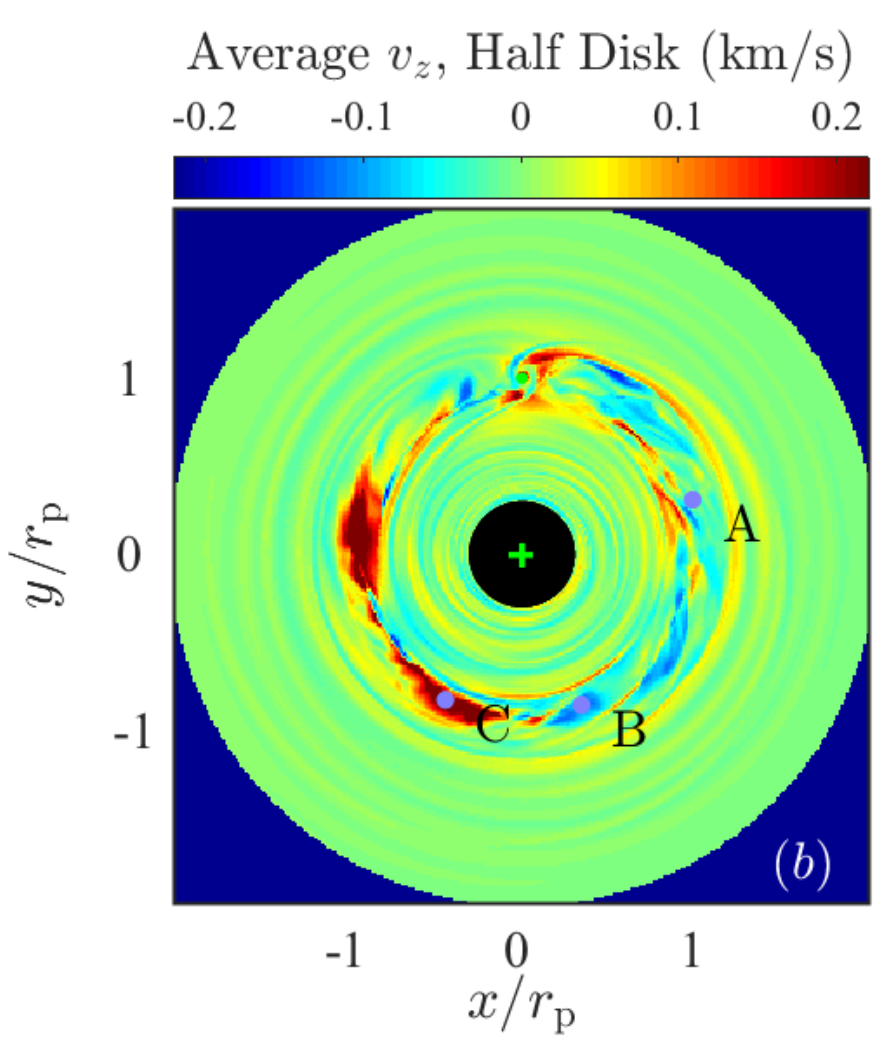}
\includegraphics[trim=0 0 0 0, clip,width=0.32\textwidth,angle=0]{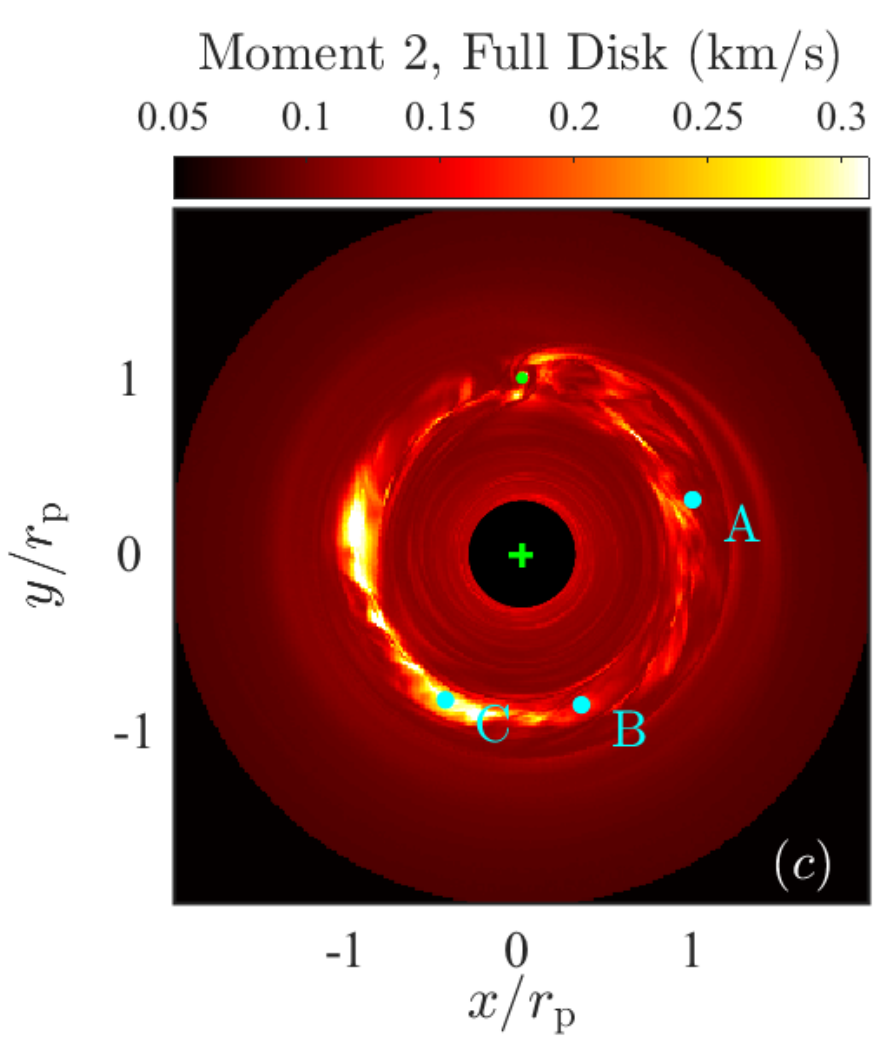}
\includegraphics[trim=0 0 0 0, clip,width=\textwidth,angle=0]{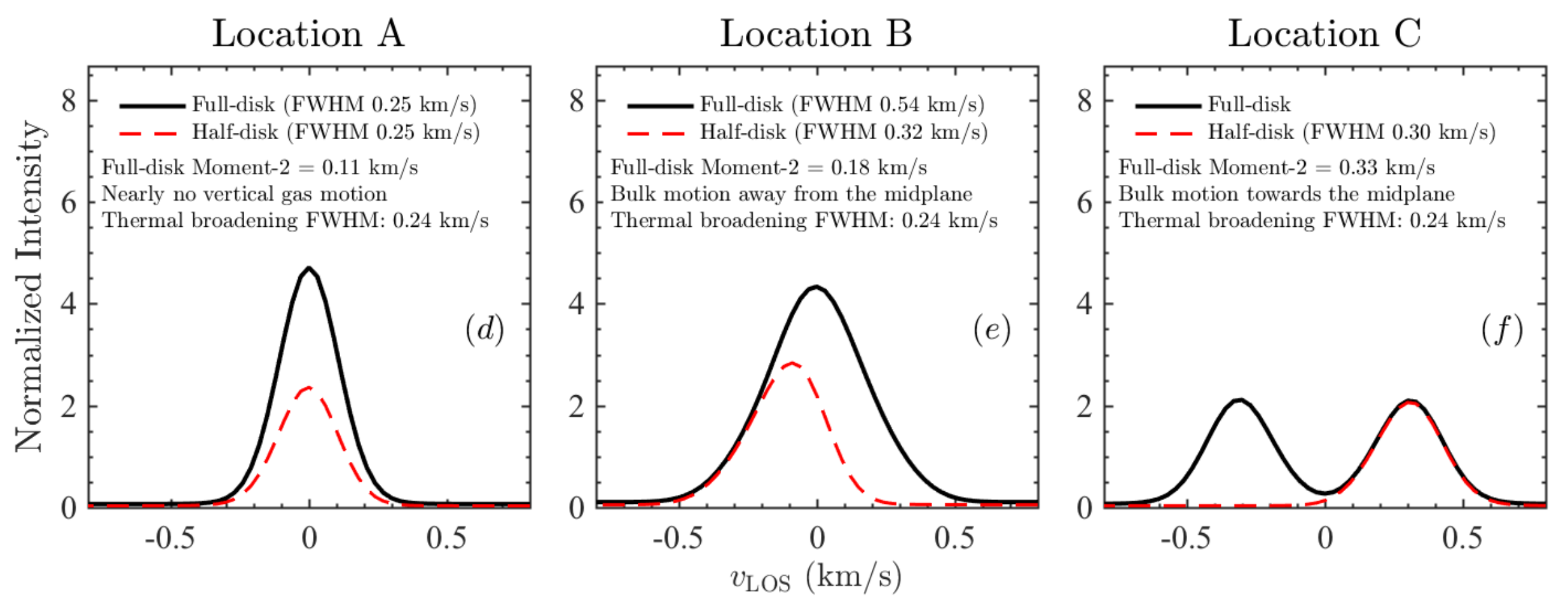}
\end{center}
\figcaption{{\it Top:} Moment-0 ($a$) and moment-2 ($c$) maps of Model Q4 in $^{13}$C$^{18}$O $J=3-2$ emission at face on, and density-weighted vertically averaged $v_z$ in the top half disk (($b$); note that the two halves are midplane-symmetric). The disk is globally optically thin. {\it Bottom:} Line profiles at Locations A, B, and C marked in the top panels (the system LOS velocity is 0); both full (solid line) and half-disk (dashed line) profiles are shown. The FWHM of a thermally broadened line at 40 K (roughly $\tmid$) is 0.24 km/s. At Location A, both profiles are consistent with thermal broadening, indicating no planet-induced motions. At Location B, the half-disk profile is non-thermally broadened to FWHM=0.32 km/s, indicating the presence of local (intra-half-disk) vertical motions; the full disk profile is further broadened to FWHM=0.54 km/s, indicating the presence of global (entire-half-disk) bulk motions. Both motions are transonic. At Location C, the global bulk motions are so large that the full disk profile is double-peaked. The sign of the peak $v_{\rm LOS}$ in the half-disk profiles indicates a bulk motion towards the midplane at Location C, and away from the midplane at Location B. See \S\ref{sec:basic} for details.
\label{fig:lineprofile_gap}}
\end{figure*}

Figure~\ref{fig:lineprofile_gap} further illustrates the origin of the planet-induced line broadening. We plot the line profiles at three locations inside the gap. The top row shows the moment-0 ($a$) and moment-2 ($c$) maps, as well as the density-weighted vertically averaged $v_z$ in the upper (closer to observers) half disk (($b$); the full disk has zero average $v_z$ everywhere due to its symmetry).

Locations A, B, and C represent three situations with different flow patterns. The line profile of optically thin gas at temperature $T$ follows a 1D  Maxwell-Boltzmann distribution
\begin{equation}
\label{eq:thermalprofile}
I(v_{\rm LOS})\propto e^{-\frac{m_\mu v_{\rm LOS}^2}{2kT}}.
\end{equation}
For gas at $T=\tmid\approx40$K inside the gap, the full-width-half-magnitude (FWHM) of a thermally broadened line is FWHM$_{\rm thermal}=2\sqrt{2\ln{2}\ k\tmid/m_\mu}=0.24$ km/s. At Location A, the FWHM is 0.25 km/s in both the half-disk and full disk models (($d$), suggesting an absence of both velocity dispersion within the half-disk (intra half-disk) and bulk velocity of the entire half-disk column (also note the zero average $v_z$ in the half-disk).

The vertical motion at Location C is completely different ($f$). First, the half-disk model line profile has a FWHM of 0.30 km/s, substantially larger than FWHM$_{\rm thermal}=0.24$ km/s, indicating the presence of intra-half-disk velocity dispersions. Secondly, the full-disk line profile is double peaked at $\pm0.3$ km/s, with each peak tracing the bulk motion in one half of the disk (the half-disk profile overlaps with the left peak in the full-disk profile). This indicates that the bulk gas in each half disk is moving at a supersonic bulk velocity of 0.30 km/s $\sim3\cs$. Both factors contribute to the resulting full-disk moment-2, 0.33 km/s, significantly larger than that at position A, 0.11 km/s. The sign of the peak $v_{\rm LOS}$ in the half-disk profile indicates a bulk motion towards the midplane in the two half disks.

The flow pattern at Location B is in between. The half disk line profile is kinematically broadened to FWHM =  0.32 km/s, comparable to Location A, but peaks at a lower $|v_{\rm LOS}|\sim0.1$ km/s, indicating a smaller global (entire half-disk) bulk motion. As a result, the full disk line profile remains single peaked, with its moment-2 (0.18 km/s) in between Locations A and C. Note that at Location B the bulk gas motion is away from the midplane, indicated by the negative peak $v_{\rm LOS}$ in the half-disk model; this is different from Location C.

\section{Discussions}\label{sec:discussions}

\subsection{Dependences of the Key Signature on Various Parameters}\label{sec:variations}

\begin{figure*}
\begin{center}
\text{\large Model Q4, Moment-2, $^{13}$C$^{18}$O, Face On (reference: $\cs(40\rm K)=0.10$ km/s)}
\includegraphics[trim=0 0 0 0, clip,width=\textwidth,angle=0]{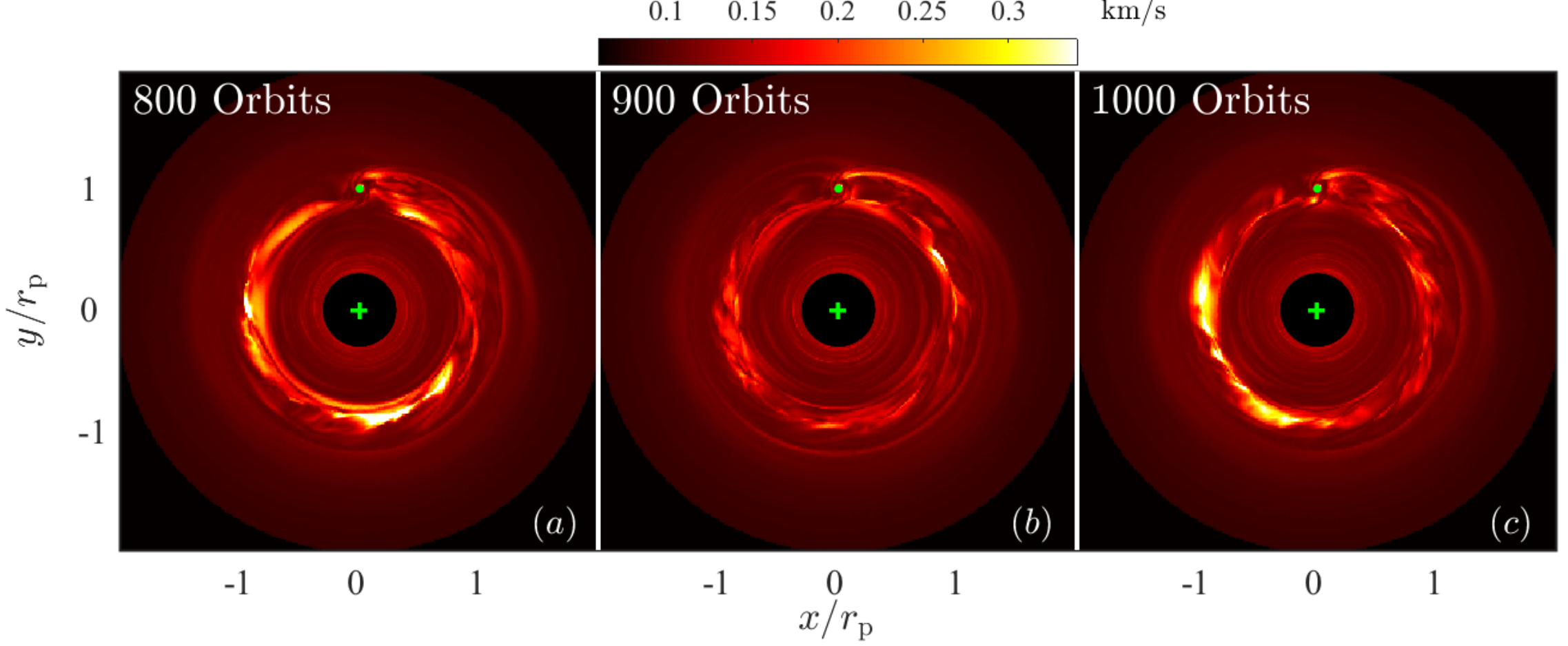}
\includegraphics[trim=0 0 0 0, clip,width=0.5\textwidth,angle=0]{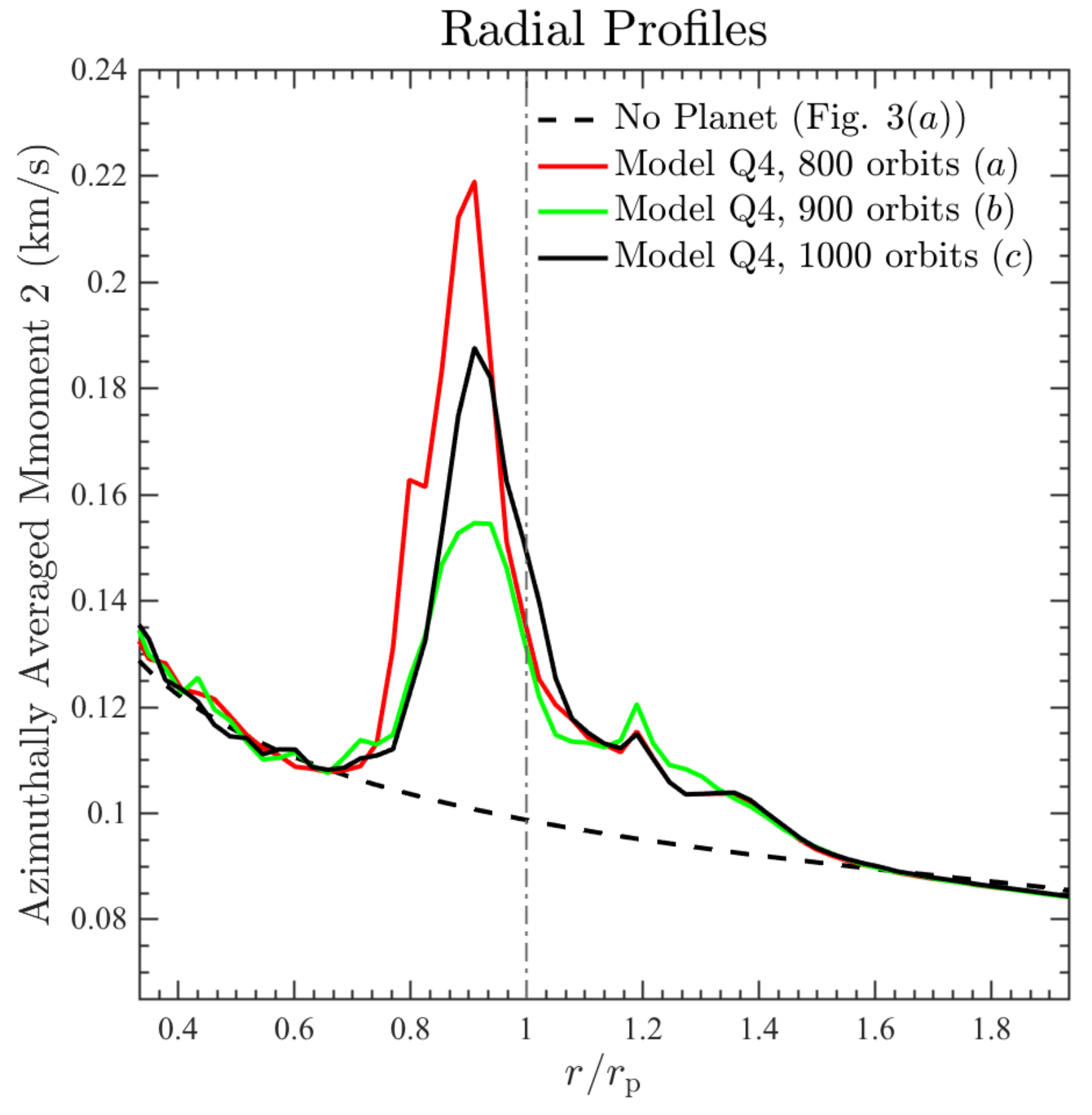}
\end{center}
\figcaption{Moment-2 in Model Q4 at 800, 900, and 1000 planetary orbits. Planet-induced turbulence fluctuates in time, while maintaining an amplitude in non-thermal broadening comparable to thermal. See \S\ref{sec:basic} for discussions.
\label{fig:evolution}}
\end{figure*}

\paragraph{Temporal Fluctuation} Figure~\ref{fig:evolution} shows moment-2 maps in Model Q4 at 800, 900, and 1000 planetary orbits. Vertical velocity dispersion at different regions inside the gap fluctuates in a seemingly uncorrelated manner. The key signature --- the ring of high velocity dispersion --- persists, while fluctuating with an azimuthally averaged peak-to-peak amplitude of $\sim$30\% (bottom panel). Note that the change is not monotonic --- the bump is lowest at 900 orbits. 
This temporal fluctuation accords with the interpretation that we find planet-induced turbulence.

\begin{figure*}
\begin{center}
\text{\large Model Q1, Moment-2, $^{13}$C$^{18}$O, Face On (reference: $\cs(40\rm K)=0.10$ km/s)}
\par\smallskip
\includegraphics[trim=0 0 0 0, clip,width=\textwidth,angle=0]{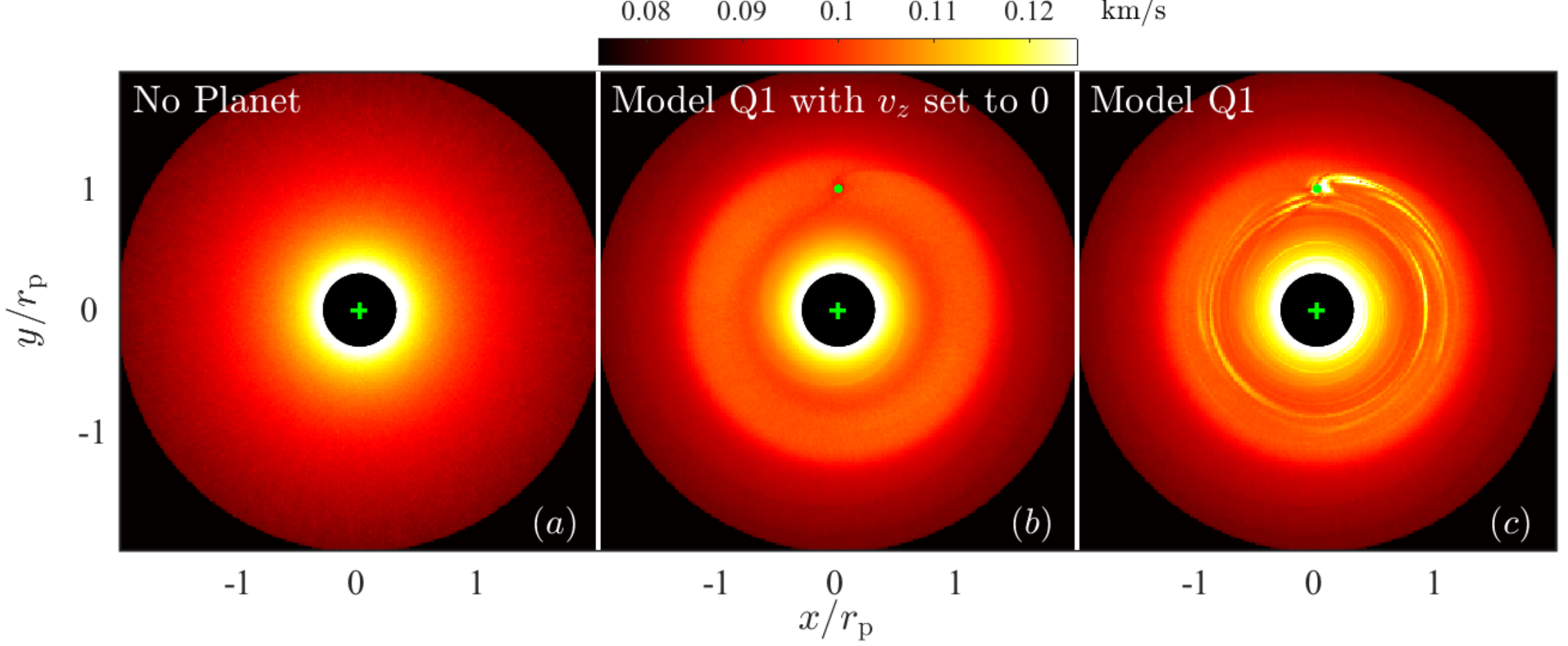}
\includegraphics[trim=0 0 0 0, clip,width=0.5\textwidth,angle=0]{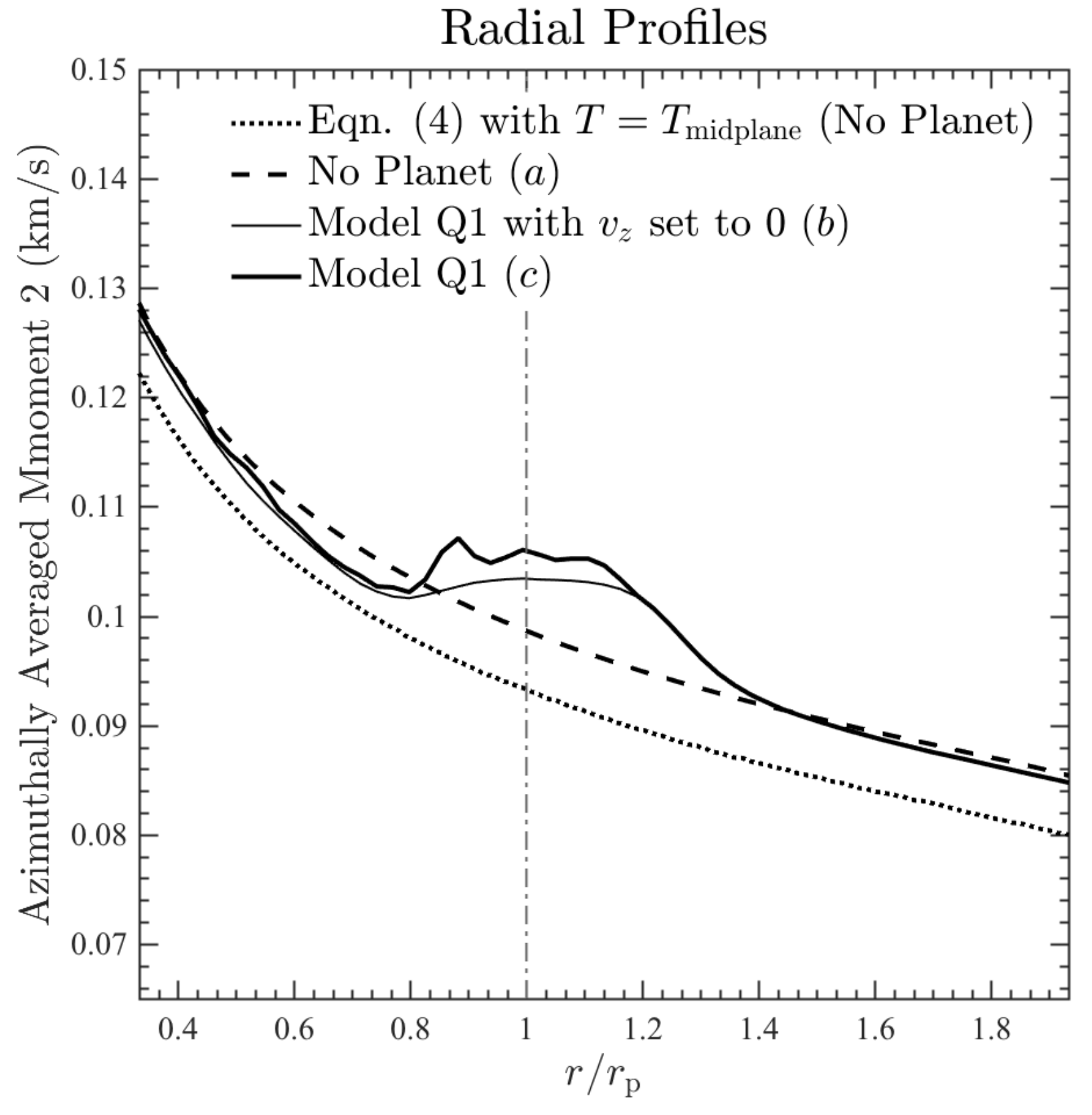}
\end{center}
\figcaption{Similar to Figure~\ref{fig:moment2_basic}, but for Model Q1 ($\mplanet=10^{-3}M_\star$). Planet-induced non-thermal line broadening (the difference between the thick and thin radial profiles) falls to a level of a few percent $\cs$ once azimuthally averaged. See \S\ref{sec:variations} for discussions.
\label{fig:moment2_1mj}}
\end{figure*}

\paragraph{Planet Mass} Figure~\ref{fig:moment2_1mj} shows moment-2 in Model Q1. The key signature drops dramatically, indicating a transition from turbulent to laminar flow when planet mass decreases from Q4 to Q1. The non-thermal velocity dispersion, once azimuthally averaged, is at the level of $\sim$10\% inside the gap. Narrow streamers of high velocity dispersion, mainly associated with the inner gap edge and spiral arms, are visible ($c$). However, extremely high angular resolutions are likely needed to resolve them in real observations.
 
\begin{figure*}
\begin{center}
\text{\large Model Q4, Moment-2, $^{13}$C$^{18}$O, Face On (reference: $\cs(40\rm K)=0.10$ km/s)} \par\smallskip
\includegraphics[trim=0 0 0 0, clip,width=0.62\textwidth,angle=0]{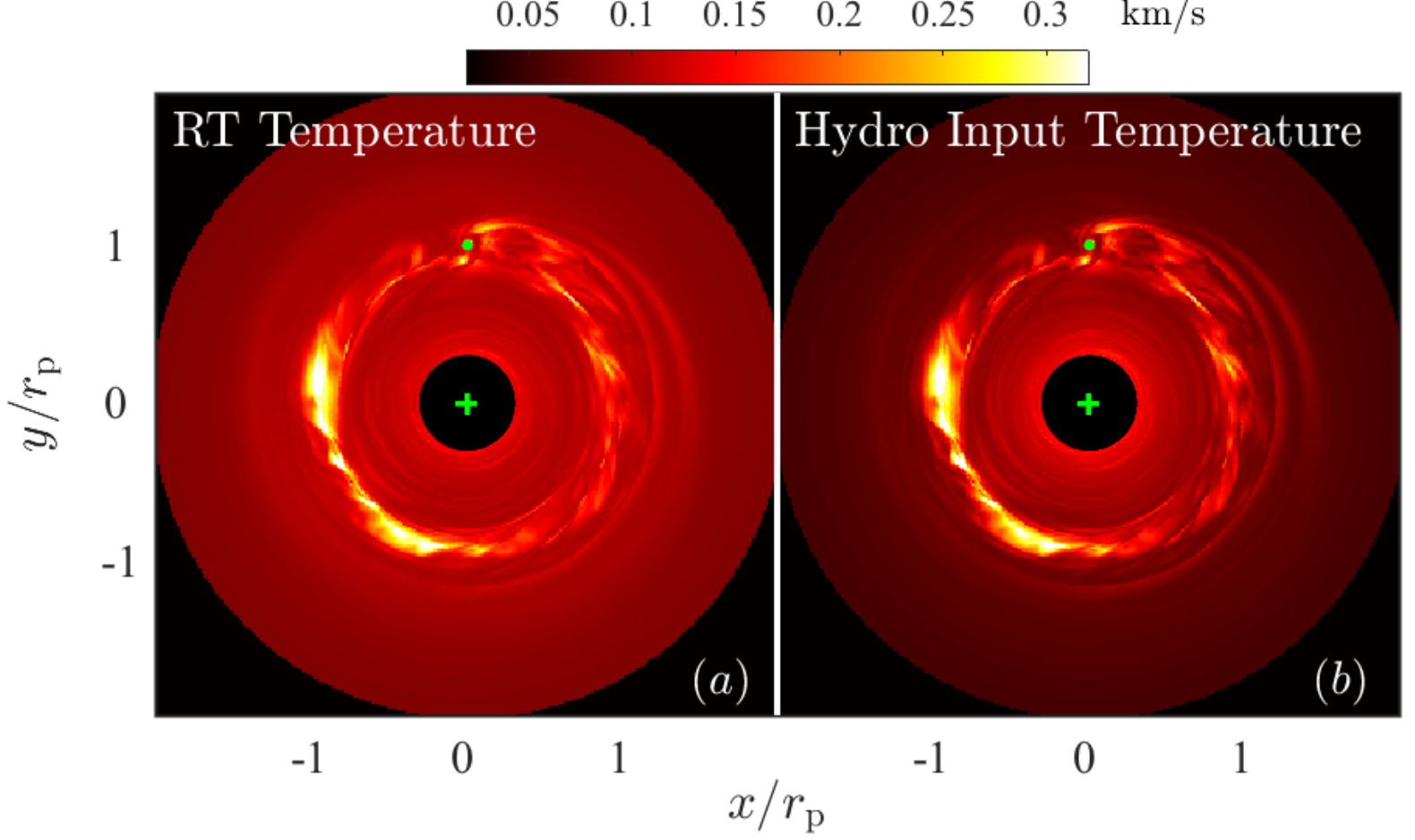}
\includegraphics[trim=0 0 0 0, clip,width=0.36\textwidth,angle=0]{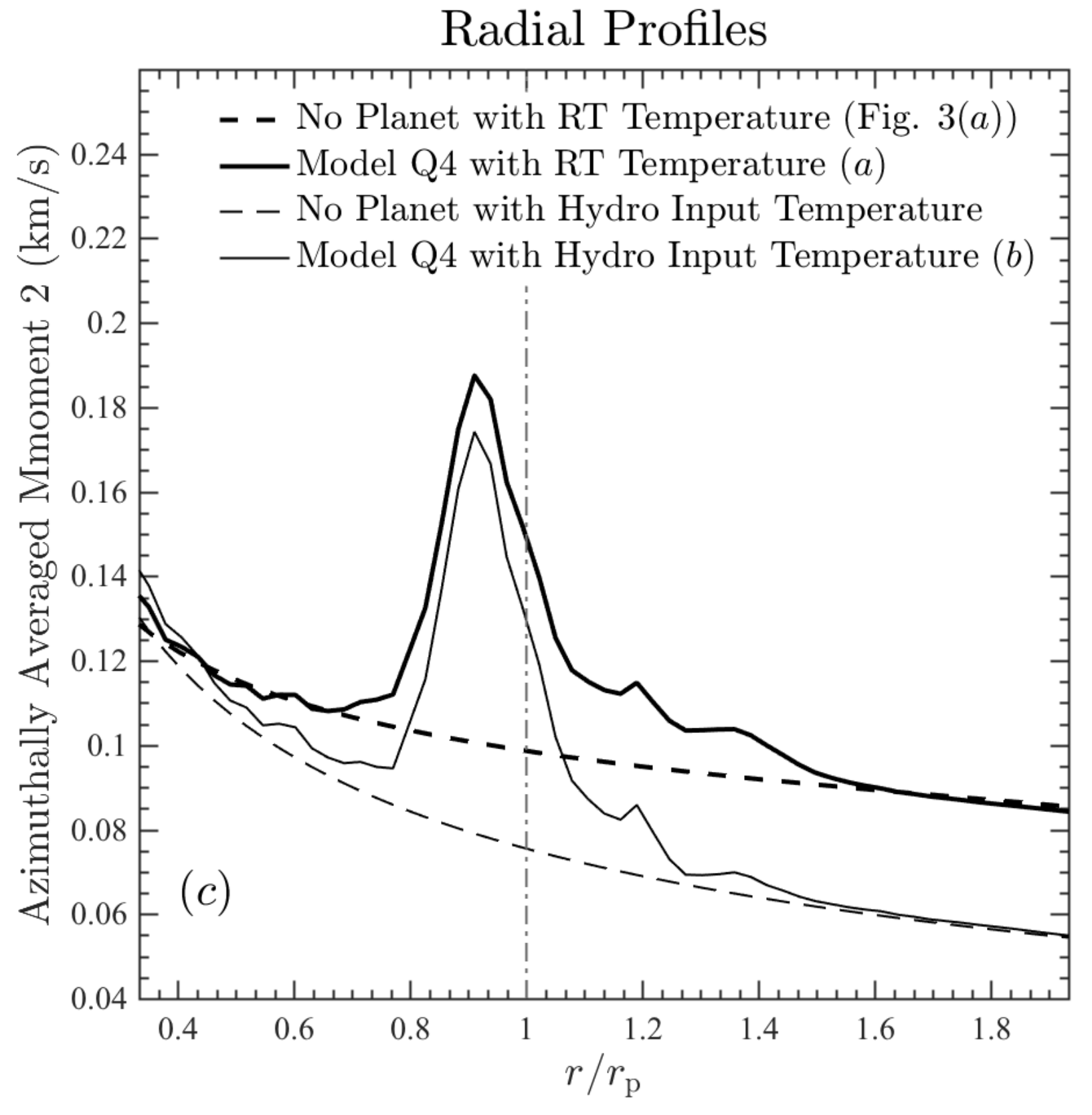}
\end{center}
\figcaption{Panel ($a$): Moment-2 in Model Q4 produced with radiative transfer (RT) temperature (the same map as in Figures~\ref{fig:moment2_basic}($c$), \ref{fig:lineprofile_gap}($c$), and \ref{fig:evolution}($c$)). Panel ($b$): Moment-2 map of the same hydro model, but produced with the temperature assumed in the hydro simulation (vertically isothermal and $h/r=0.05$). Panel ($c$) shows their azimuthally averaged radial profiles, as well as the profiles in the no planet cases. The key planet-induced signature --- the high velocity dispersion inside the gap (the difference between the solid and dashed line in each set) --- stays robust. See \S\ref{sec:variations} for discussions.
\label{fig:hydrot}}
\end{figure*}

\paragraph{Temperature Structure} 
As mentioned in \S\ref{sec:rt}, density and temperature in our models are not self-consistent. Here we carry out an experiment to qualitatively assess its effect on the key signature. Figure~\ref{fig:hydrot} compares two moment-2 maps of Model Q4, one with radiative transfer temperature adopted in the line emission simulation ($a$), and the other with the temperature corresponding to Equation \ref{eq:hydro_cs} (hydro input temperature; ($b$)). Different temperatures result in different velocity dispersions in the thermally broadened inner and outer disks. Inside the gap, however, the two converge as kinematic broadening dominates, and the key signature remains robust.

\begin{figure*}
\begin{center}
\text{\large Model Q4, Moment-2, CO, Face On (reference: $\cs(40\rm K)=0.10$ km/s)} \par\smallskip
\includegraphics[trim=0 0 0 0, clip,width=0.63\textwidth,angle=0]{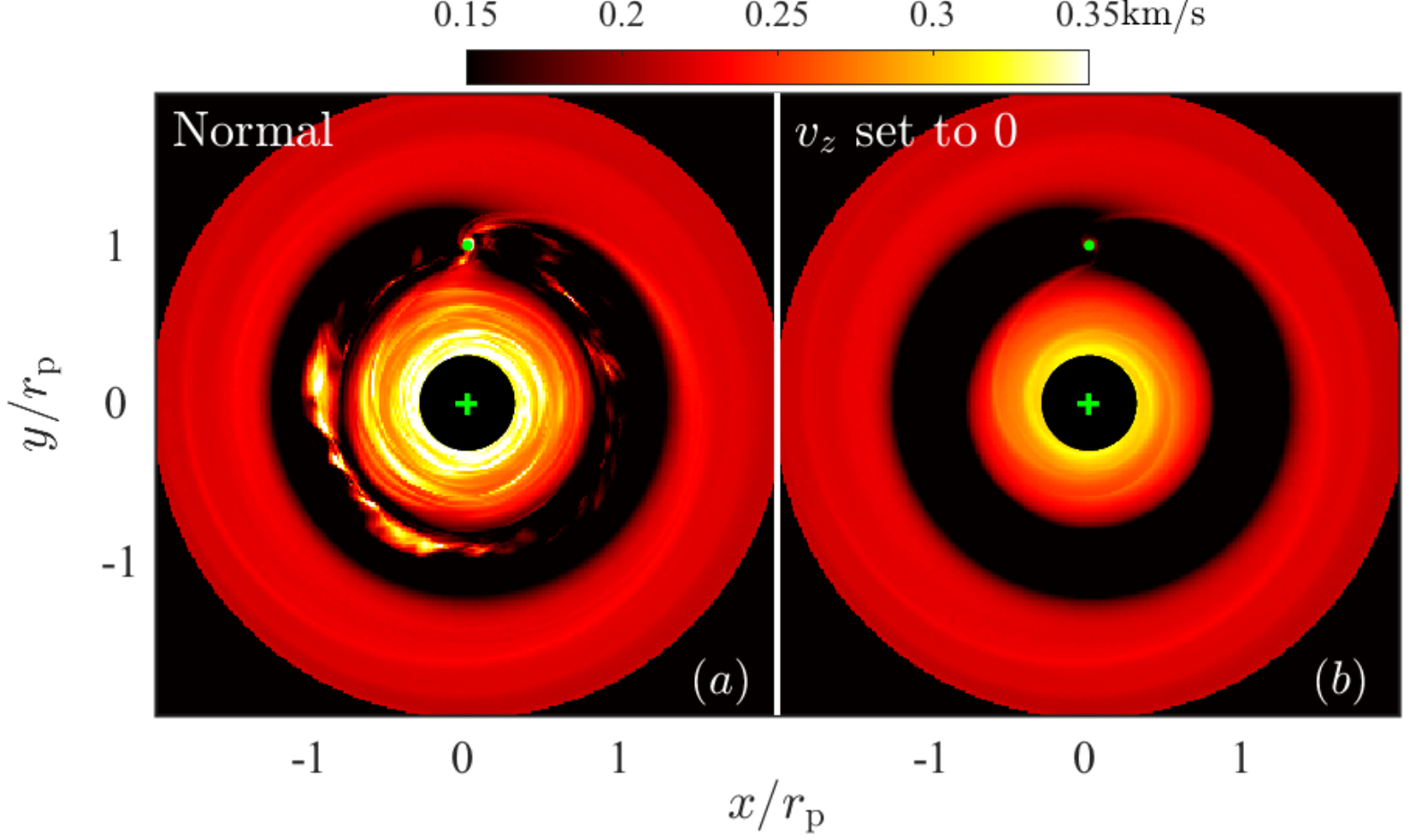}
\includegraphics[trim=0 0 0 0, clip,width=0.36\textwidth,angle=0]{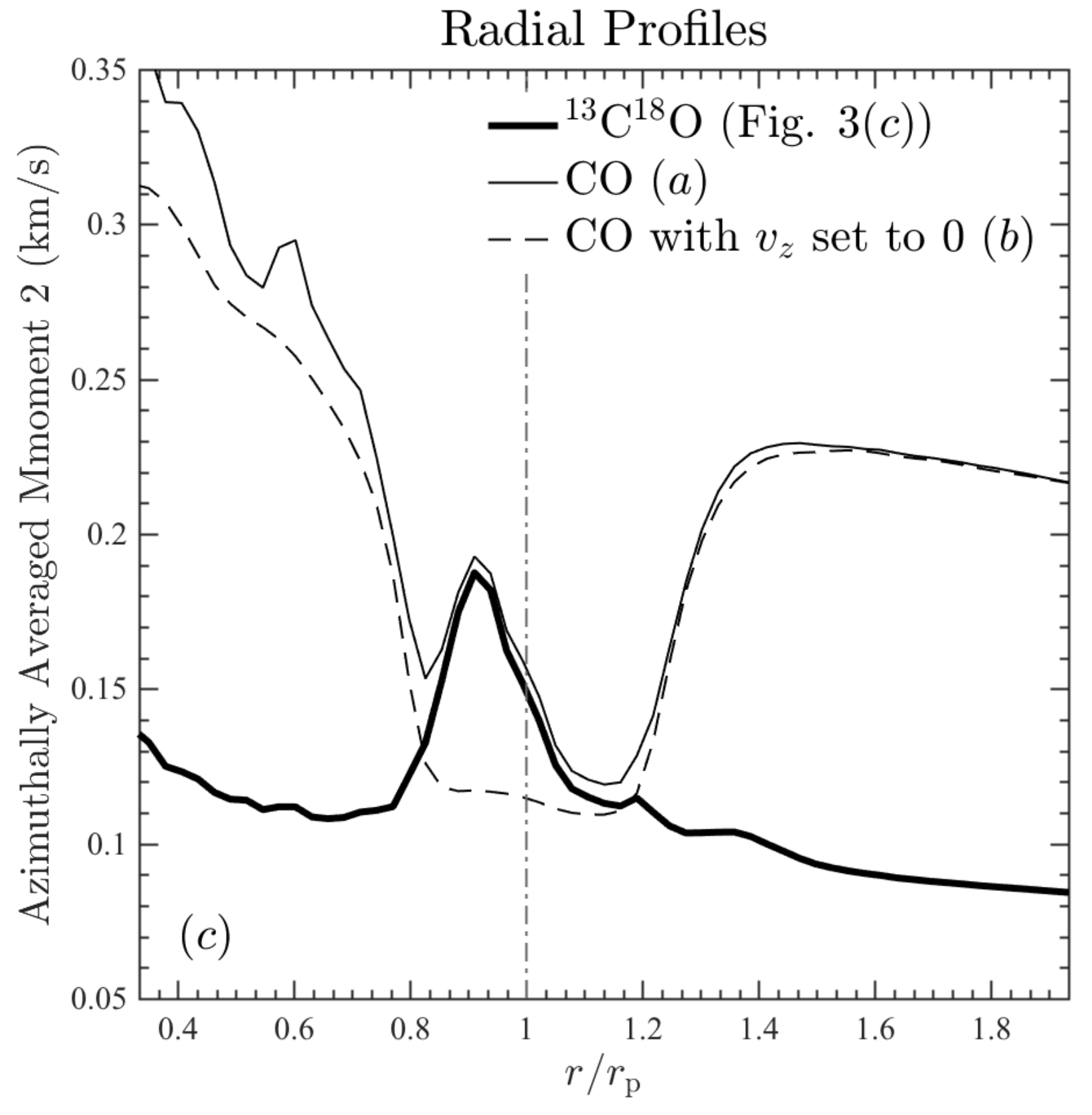}
\end{center}
\figcaption{Moment-2 in Model Q4 in CO ($a$), the same map produced with the vertical velocity $v_z$ manually set to 0 ($b$), and their azimuthally averaged radial profiles ($c$). The inner and outer disks are extremely optically thick to CO, which traces warm layers above the midplane; consequently, thermal broadening is larger than in the $^{13}$C$^{18}$O case. The gap region remains optically thin. The narrow arcs and rings with high velocity dispersions in the inner disk in ($a$) are absent in ($b$). They are induced by the planet in the surface layer. The key planet-induced signature in velocity dispersion inside the gap, which is the difference between the thin solid and dashed curves in ($c$), stays robust. Note that the thick black curve in the radial profile plot in Figures~\ref{fig:moment2_basic}, \ref{fig:evolution}, \ref{fig:hydrot}, and here is the same. See \S\ref{sec:variations} for discussions.
\label{fig:moment2_co}}
\end{figure*}

\paragraph{Tenuous Gas Tracers}
Figure~\ref{fig:moment2_co} compares moment-2 in Model Q4 in CO ($a$) and in a control case produced with $v_z$ manually set to 0 ($b$). While the gap is still optically thin to CO emission ($\tau$ peaks at $\sim$0.01 at $v_{\rm LOS}=0$), the inner and outer disks are extremely optically thick ($\tau\gtrsim1000$). Therefore outside the gap CO traces a hot surface layer above the midplane, leading to larger thermal broadening than in the midplane-tracing $^{13}$C$^{18}$O emission. Nevertheless, The regions with the highest velocity dispersion inside the gap are still visible in the 2D map ($a$), and the key planet-induced signature -- the difference between the thin solid and dashed curves in ($c$) --- is still prominent. Furthermore, the high velocity dispersions seen in a few narrow arcs and rings in the inner disk in ($a$), absent in ($b$), are caused by planet-induced vertical motions. While the planet does not induce bulk motions in the inner disk, it does induce vertical motions in the surface layer, visible only in tenuous gas tracers like CO. 

\begin{figure*}
\begin{center}
\text{\large Model Q4, Moment-2 (reference: $\cs(40\rm K)=0.10$ km/s)} \par\smallskip
\text{\large $^{13}$C$^{18}$O\ \ \ \ \ \ \ \ \ \ \ \ \ \ \ \ \ \ \ \ \ \ \ \ \ \ \ \ \ \ \ \ \ \ \ \ \ \ \ \ \ \ \ \ \ \ \ \ \ \ \ \ \ \ \ \ \ \ \ CO} \par\smallskip
\includegraphics[trim=0 0 0 0, clip,width=0.49\textwidth,angle=0]{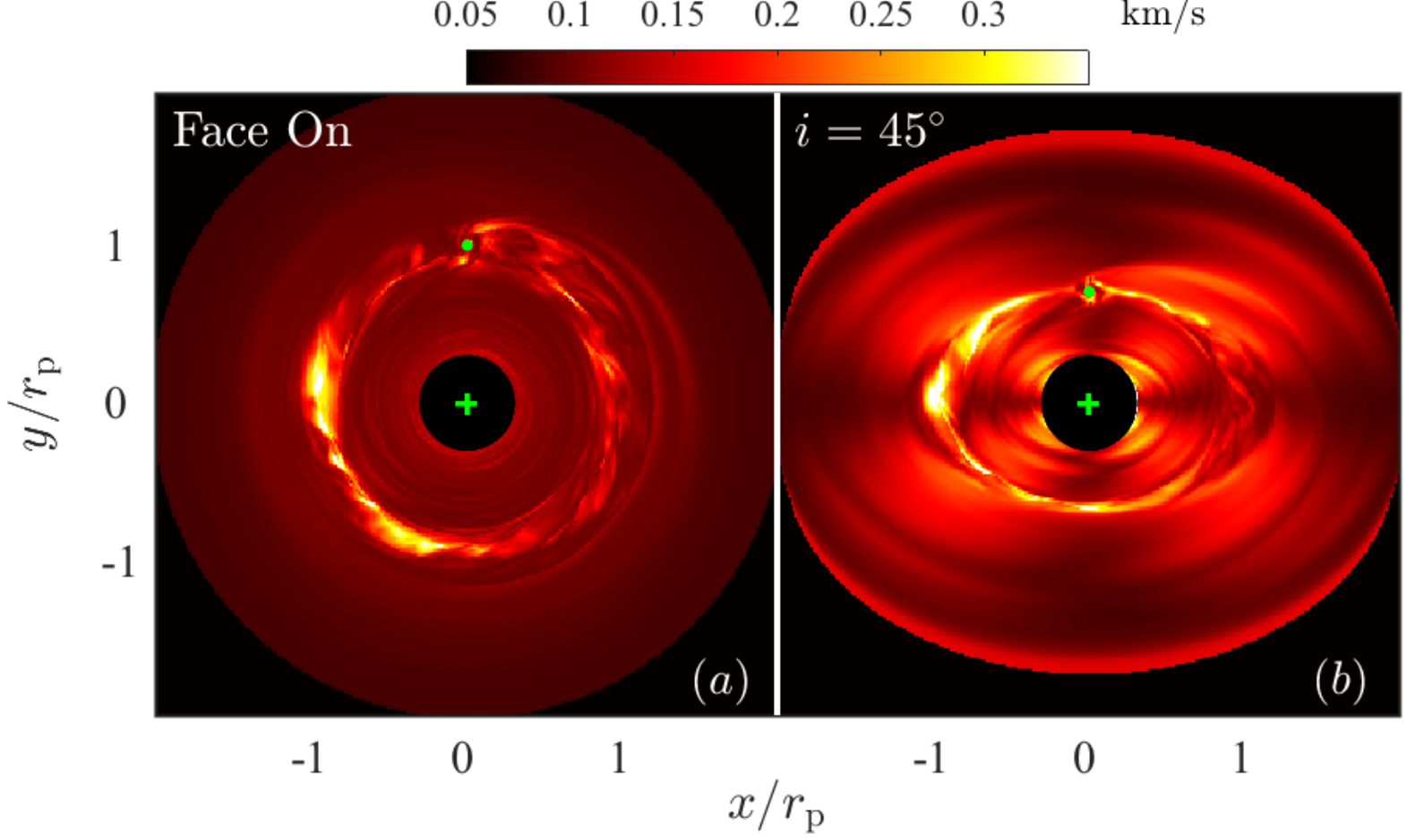}
\includegraphics[trim=0 0 0 0, clip,width=0.49\textwidth,angle=0]{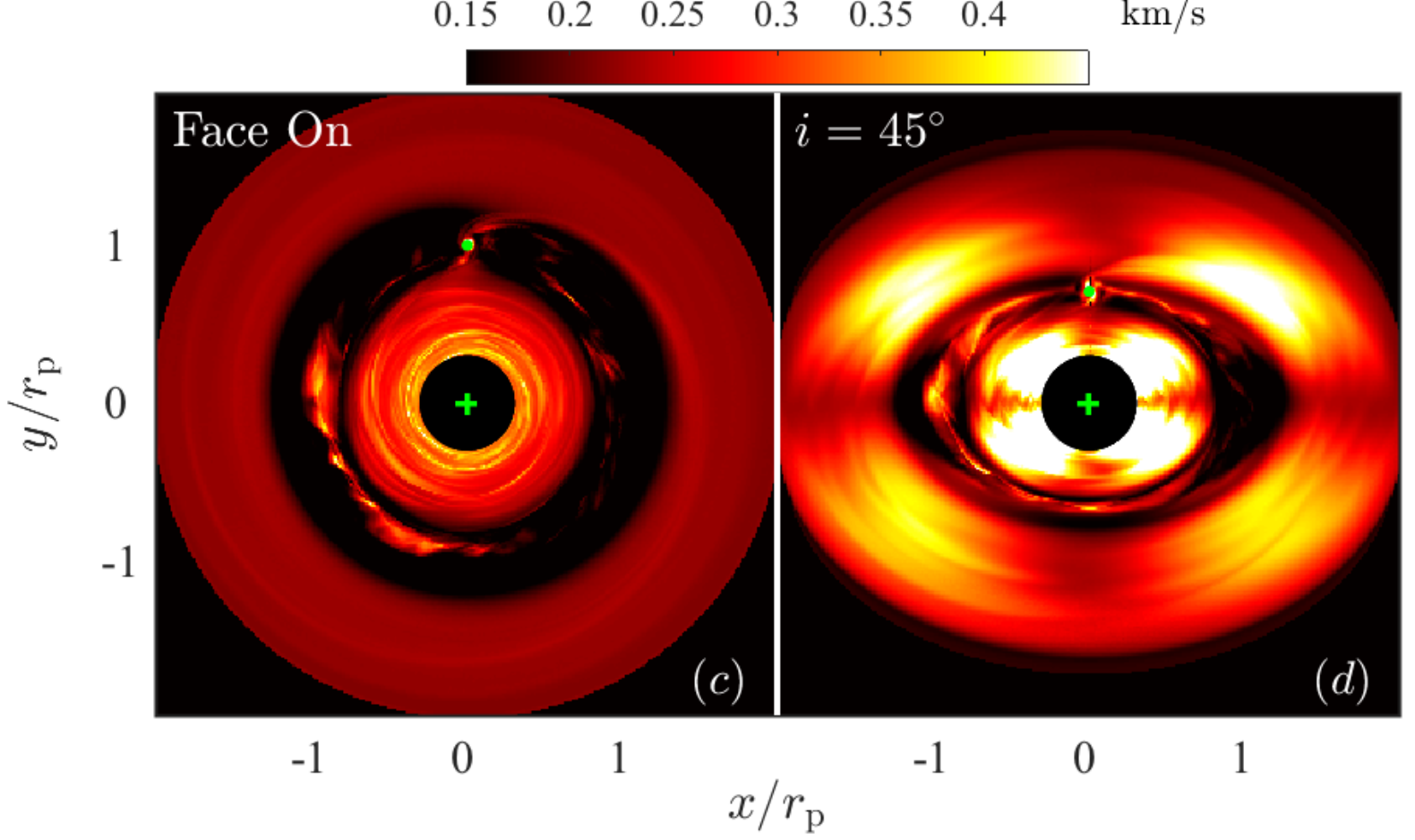}
\end{center}
\figcaption{Moment-2 in Model Q4 in $^{13}$C$^{18}$O ($a,b$) and CO ($c,d$) at face-on ($a,c$) and inclination $i=45^\circ$ ($b,d$). The northern part is on the far side and the major axis is along East--West. The high velocity dispersions in the quadruple pattern in ($b$) and ($d$) are introduced by the differences in the projected Keplerian rotation along each LOS (Keplerian broadening). Planet-induced meridional line broadening inside the gap are visible in modestly inclined disks. See \S\ref{sec:variations} for discussions.
\label{fig:moment2_inclination}}
\end{figure*}

\paragraph{Finite Inclinations}
Figure~\ref{fig:moment2_inclination} compares moment-2 in Model Q4 at face-on and inclination $i=45^\circ$ in $^{13}$C$^{18}$O and CO. Under modest $i$, each LOS passes through regions with different heliocentric distances and effective inclinations, due to the finite disk thickness. These regions have different LOS-projected Keplerian rotation (most effectively in between the major and minor axes), which broadens emission lines (a.k.a. Keplerian broadening or Doppler broadening). Nevertheless, regions inside the gap with planet-induced high velocity dispersions are still visible. Note that in this case planet-induced radial motions contribute too.

\subsection{Planet-Induced Vertical Motions Outside the Gap}\label{sec:outsidegap}

\begin{figure*}
\begin{center}
\text{\large Model Q4, $^{13}$C$^{18}$O, Face On (reference: $\cs(40\rm K)=0.10$ km/s)}
\par\smallskip
\includegraphics[trim=0 0 0 0, clip,width=0.32\textwidth,angle=0]{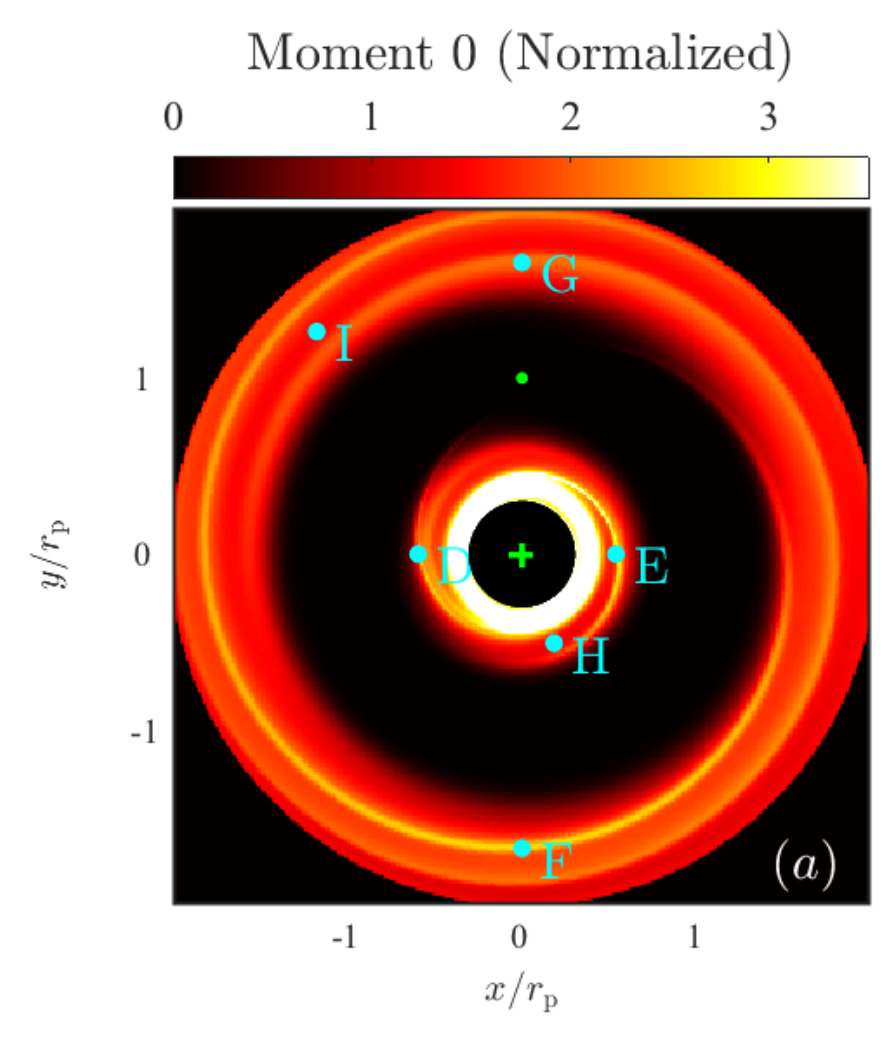}
\includegraphics[trim=0 0 0 0, clip,width=0.32\textwidth,angle=0]{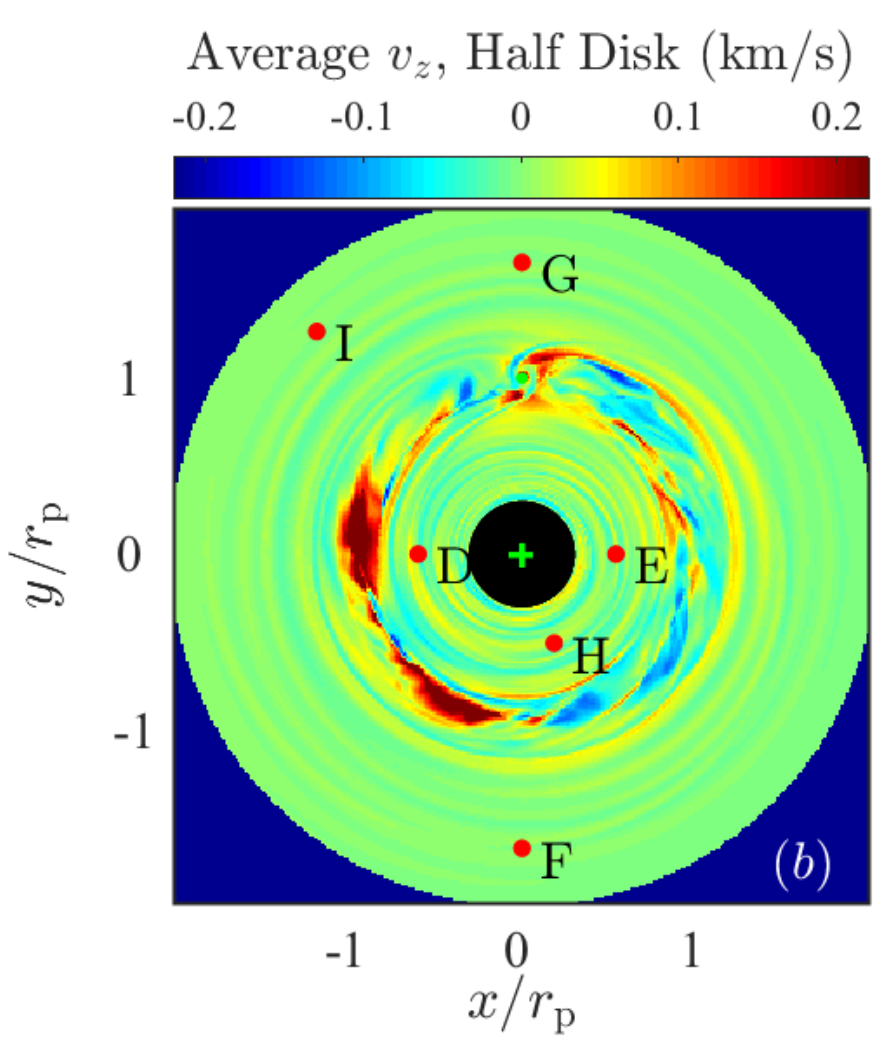}
\includegraphics[trim=0 0 0 0, clip,width=0.32\textwidth,angle=0]{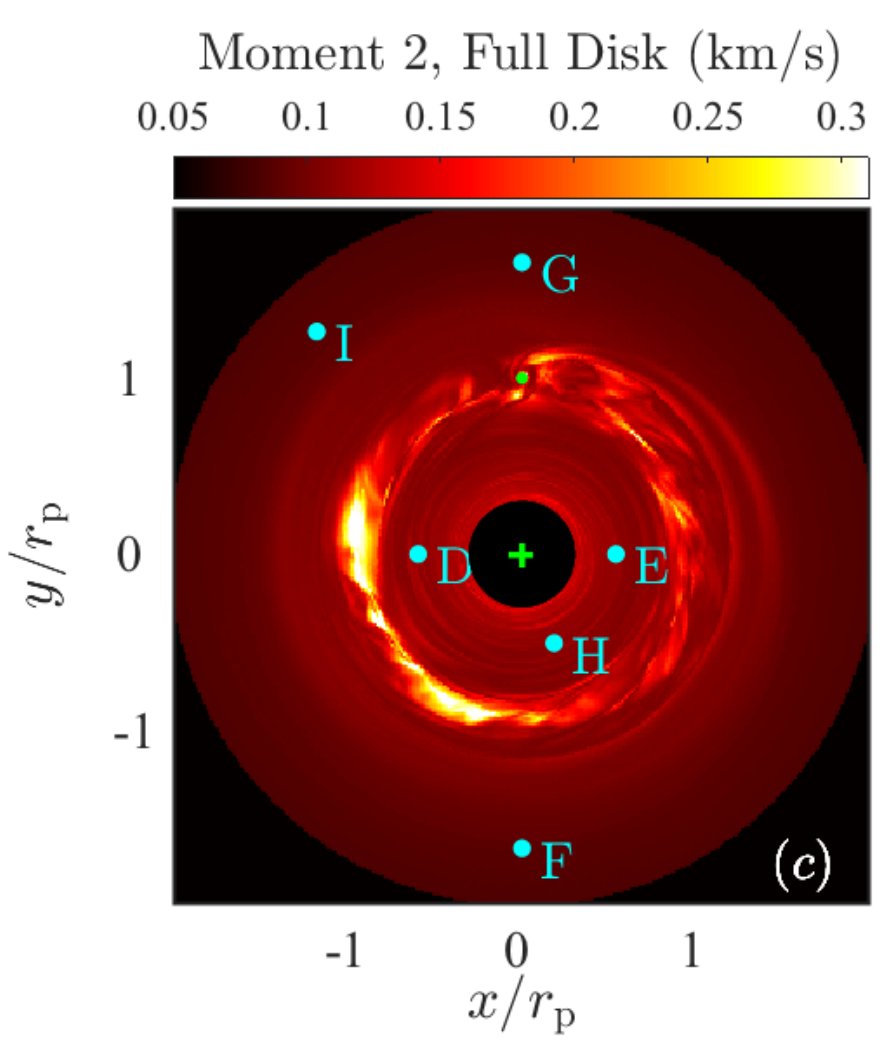}
\includegraphics[trim=0 0 0 0, clip,width=\textwidth,angle=0]{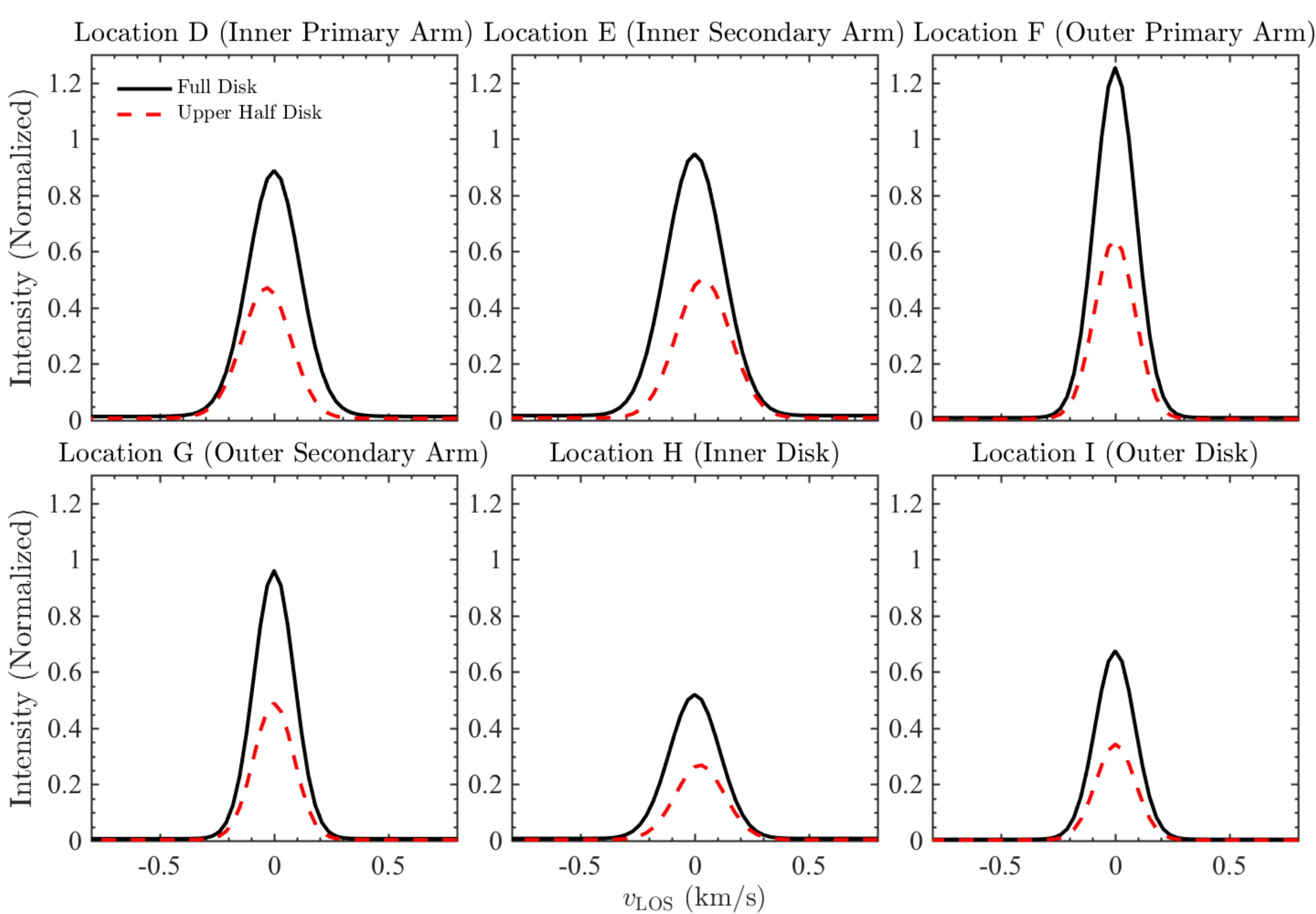}
\end{center}
\figcaption{Similar to Figure~\ref{fig:lineprofile_gap}, but for another 6 locations outside the gap: the inner primary and secondary arms (D and E), the outer primary and secondary arms (F and G), and the inter-arm regions in the inner and outer disks (H and I). The planet does not introduce bulk vertical motions at Locations F -- I, as their half-disk line profiles peak at $v_{\rm LOS}=0$. At  the two inner arms (Locations D and E), half disk line profiles peak at $|v_{\rm LOS}|\sim0.05$ km/s $\sim0.5\cs$ with opposite signs, indicating the presence of transonic bulk vertical motions away from the midplane at the primary arm and towards the midplane at the secondary arm. The line broadening in the full disk model is consistent with thermal at Locations F -- I, while additional broadening is present at the two inner arms. See \S\ref{sec:outsidegap} for discussions.
\label{fig:lineprofile_outsidegap}}
\end{figure*}

Figure~\ref{fig:lineprofile_outsidegap} shows $^{13}$C$^{18}$O line profiles at six locations outside the gap in Model Q4: the inner primary and secondary arms (D and E), the outer primary and secondary arms (F and G), and the inner and outer disks (H and I). At the latter four locations, the half-disk line profile peaks at $\vlos=0$, indicating an absence of global vertical motion with respect to the midplane. Meanwhile, the full disk line profiles are almost entirely broadened thermally (cf. comparing Figure~\ref{fig:moment2_basic}($a$) and \ref{fig:moment2_basic}($c$)), suggesting an absence of local (intra-half-disk) velocity dispersion as well. 

On the other hand, at the two inner spiral arms the half-disk line profile peaks at $|v_{\rm LOS}|\sim0.05$ km/s $\sim0.5\cs$, indicating the presence of transonic global vertical motions with respect to the midplane. Such motions have previously been seen \citep[e.g.,][]{lyra16, riols18}. Note that shock heating at the arms \citep[e.g.,][]{richert15, hord17}, expected to further modify the flow pattern, is not included. The line width in the half-disk model at both arms are consistent with thermal broadening, indicating an absence of intra-half-disk velocity dispersion. Interestingly, gas move {\it away from} the midplane at the primary arm, and {\it towards} the midplane at the secondary arm. The kinematic support at the arms helps lift the disk surface in addition to the usual vertical pressure support \citep[][Fig. 3]{dong17spiralarm}, echoing with previous findings from 3D hydro$+$RT simulations that the inner arms are more prominent than the outer arms in scattered light \citep[e.g.,][]{dong15spiralarm, zhu15densitywaves}. 

\section{Comments on Model Applications}\label{sec:applications}

Searching for planet-induced turbulence inside gaps is similar in nature to searching for other types of turbulence induced by instabilities, such as MRI (\S\ref{sec:intro}). Naturally, the techniques developed for the latter applies to the former. Here we briefly outline a generic roadmap for real life applications of our models.

When probing a planet-opened gap using a molecular line to which the gap region is optically thin, the line is broadened by four factors: (A) Keplerian broadening, (B) thermal broadening, (C) planet-induced broadening, and (D) other turbulent broadening. To infer the presence and measure the strength of (C), we need to estimate the contributions from (A), (B), and (D), and take them out of the observed line broadening.

Examples of assessing (A) and (B) using line observations can be found in, e.g., \citet{flaherty15, flaherty17, flaherty18} and \citet{teague16, teague18twhya}. To estimate thermal broadening, an accurate assessment of the temperature at the emitting regions is needed. Techniques have been developed recently to do so using observations of multiple transitions from a single molecule, and temperatures can now be constrained to within a few K in certain cases \citep[e.g., TW Hya,][]{teague18twhya}. Once the contributions from (A) and (B) are taken out, additional broadening, if present, can be attributed to (C) and (D). Separate efforts are then needed to distinguish (C) from (D). Note that so far the searches for (D) have generally returned null results, putting upper limits on $v_{\rm turb,(D)}$ at a few percent of $\cs$ (\S\ref{sec:intro}).

In order to minimize the contributions from (A) and (B), disks close to face-on are preferred, so do heavy molecule as they naturally exhibit lower thermal broadening at a given temperature (turbulent broadening is not expected to depend on molecular weight).

We emphasize again that the experiments here are designed to best illustrate the key planet-induced signature, instead of representing best practices in real life. The detectability of gas emission from a real gap mainly depends on its gas surface density (measurable; e.g., \citealt{vandermarel15, vandermarel16}), and the choice of the tracer.

\subsection{Planet-Induced Velocity Dispersion as Signposts of Planets}\label{sec:advantages}

Recent years have witnessed the discoveries of spiral arms, gaps, and dust traps in many protoplanetary disks \citep[e.g.,][]{muto12, vandermarel13, casassus13, canovas13, isella16hd163296, follette17, pohl17hd169142, stolker17, huang18, dipierro18, dong18mwc758}. Two general categories of mechanisms have been proposed for their origins: dynamical companion-disk interactions \citep[e.g.,][]{zhu11, zhu14stone, dong15spiralarm, dipierro15hltau, bae16sao206462, bae17, dong17doublegap}, or alternative mechanisms involving either disk evolution \citep[e.g.,][]{owen11, pinilla12dusttrapping, takahashi14, mittal15, dong15giarm, suriano17, wang17, ercolano18} or snowline \citep[e.g.,][]{zhang15, okuzumi16}. 

In the companion scenario, the properties of feature-driving companions, such as location, orbit, and mass, may be constrained using observed feature morphology \citep[e.g.,][]{fung15, dong17gap, dong17spiralarm}. This provides an extremely valuable channel in using disk observations to test planet formation models. Unfortunately, planetary and non-planetary mechanisms sometimes produce similar observational features \citep[e.g.,][]{dong18spiral}. Except in rare cases \citep[e.g., the stellar mass companion in HD 100453,][]{dong16hd100453, wagner18}, it is difficult to confirm the planetary origin of observed features, and subsequently to take advantages of disk observations to study planet formation.

Planet-induced kinematics signatures provide another possible avenue to help distinguish planetary and non-planetary mechanisms (\S\ref{sec:intro}). The detection of high velocity dispersions inside gaps, as studied here, can potentially lend strong support to the planet scenario. In addition, the strength in these non-thermal dispersions may be used to constrain the planet mass.

\begin{figure*}
\begin{center}
\text{\large Model Q4, $\rp=30$ AU, Moment-1, $i=45^\circ$} \par\smallskip
\includegraphics[trim=0 0 0 0, clip,width=0.75\textwidth,angle=0]{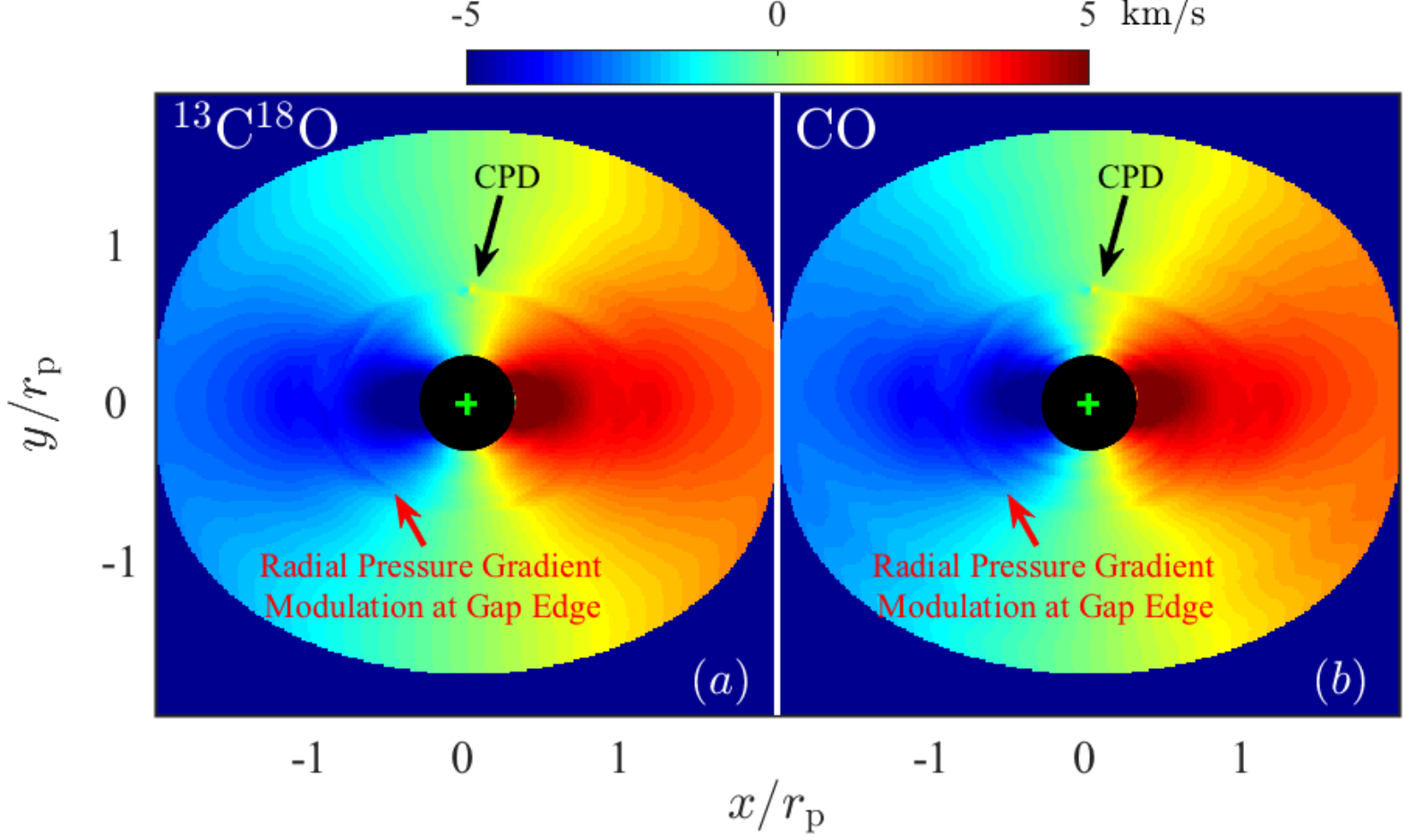}
\end{center}
\figcaption{Moment-1 maps for Model Q4 in $^{13}$C$^{18}$O and CO at 45$^\circ$ inclination. The kinematic CPD feature and the small modulations of the rotational velocity induced by the variations in the radial pressure gradient are labeled. See \S\ref{sec:advantages} for discussions.
\label{fig:moment1_i45}}
\end{figure*}

Quite a few types of planet-induced kinematics signatures have been explored (\S\ref{sec:intro}). Comparing with them, the signature studied here has four characteristics. First, it is a global signature on scales comparable to the gap width, typically 25\%--50\%$\rp$, easing the requirement on angular resolution. Further, its signal-to-noise ratio can be boosted by azimuthal averaging. Certain types of planet-induced kinematics, such as the CPD rotation \citep{perez15}, are local features detectable only under high angular resolutions. Figure~\ref{fig:moment1_i45} shows the moment-1 maps of Model Q4 in both $^{13}$C$^{18}$O ($a$) and CO ($b$) at $i=45^\circ$. The CPD rotation, clearly visible, is confined to a region comparable to the planet's Hill sphere, which is $\sim$10 times smaller than $\rp$ for Jovian planets.

Secondly, velocity dispersion in the gap driven by the planet is arguably a more numerically robust feature than other planet-induced kinematic signatures such as the CPD and local kinks in the isovelocity maps. The moment-2 signals reported here have length scales larger than local scales such as the planet's Hill radius or disk scale height, making it easier to be resolved and captured in hydrodynamics simulations.

Thirdly, the requirement on the accuracy of gap temperature assessment is lenient. Turbulent motions can be mildly supersonic, and the total line broadening can double that of thermal. Even in disks with modest inclinations and substantial Keplerian broadening, the planet-induced signature is still easily visible (Figure~\ref{fig:moment2_inclination}). Knowing the local temperature to within a factor of 2 is likely sufficient to reveal velocity dispersions induced by a multi-$\mj$ planet. This is in contrast with certain types of planet-induced kinematics. For example, \citet{teague18} showed that the variations in the gas rotations due to the modulations in the radial pressure gradient induced by Jovian planets (the weak radial structures at close to $\rp$ in Figure~\ref{fig:moment1_i45}) may only be at the level of 1\% $\vk$, or perhaps 10\% $\cs$. Measuring such signals demand accurate assessments of disk temperatures to a few percent.

Finally, the bottleneck in real-life application is expected to be the high sensitivity required to probe the weak line emission from deeply depleted gaps. With the current Atacama Large Millimeter/submillimeter Array (ALMA), this poses a strong challenge, if possible at all, for gaps at a few tens of AU, while the situation improves dramatically for giant planets at large radii. For example, if we observe Model Q4 at 140 pc with an angular resolution of 0\arcsec.1 ($0.5\rp$) to just resolve the gap, we need up to a few weeks of ALMA integration time to achieve a 10$\sigma$ detection of a CO emission line with a spectral resolution of 0.1 km/s inside a $\tau_{\rm CO}=0.1$ gap (i.e., the scenario in Figure~\ref{fig:moment2_co}). The situation will be much more favorable if the gap is at a larger radius --- should we place the planet at 100 AU with everything else being equal, an angular resolution of 0\arcsec.3 would be sufficient to resolve the gap, and the integration time for the same 10$\sigma$ detection would be reduced to a few hours.


\section{Summary}\label{sec:summary}

In this paper we use 3D hydrodynamics and radiative transfer simulations to study the signatures of planet-induced turbulence in gaps in gas line observations. A planet with $\mplanet=4\times10^{-3}M_\star$ (4 $\mj$ around a 1 $\msun$ star) can induce transonic to mildly supersonic turbulent motions inside its gap, on both local (intra-half-disk) and global (entire half-disk) scales (Figure~\ref{fig:lineprofile_gap}). The resulting velocity dispersion, after azimuthal averaging, can double that of thermal (Figure~\ref{fig:moment2_basic}). The turbulent motion can result in double-peaked line profiles at certain regions in the gap (Figure~\ref{fig:lineprofile_gap}). 
This basic key signature is robust against temporal variations (Figure~\ref{fig:evolution}), different treatments of disk temperature (Figure~\ref{fig:hydrot}), finite inclinations (Figure~\ref{fig:moment2_inclination}), and is visible in tracers with a variety of opacity as long as the gap is optically thin (Figure~\ref{fig:moment2_co}). For a planet with $\mplanet=10^{-3}M_\star$, azimuthally averaged turbulent velocity dispersion drops to $\sim$10\%$\cs$ (Figure~\ref{fig:moment2_1mj}). 
In the dense outer disk, the planet does not introduce significant vertical motions; this is true in the inner disk too, except at the surface (Figure~\ref{fig:moment2_co}) and in the spiral arms (Figure~\ref{fig:lineprofile_outsidegap}; a bulk motion away from the midplane in the primary arm, and opposite in the secondary arm).

Detecting planet-induced velocity dispersion can lend support to the planetary origin of gaps, and its strength may potentially be used for dynamical planet mass assessments. Comparing with other types of planet-induced kinematic signatures, velocity dispersion in gaps has its own characteristics. It is on radial scales comparable to the gap width, and visible after azimuthal averaging. Searching for this signature likely requires an assessment of the gap gas temperature to within a factor of 2, meanwhile a high sensitivity is needed to detect the weak line emission from deep gaps.


\section*{Acknowledgments}

We thank the anonymous referee for constructive suggestions, and Xue-ning Bai for discussions. R.D. acknowledges the support from the Natural Sciences and Engineering Research Council of Canada. Part of this work was performed at the Aspen Center for Physics, supported by US NSF grant PHY-1607611. 

\software{
\texttt{PEnGUIn} \citep{fung15thesis},
\texttt{HOCHUNK3D} \citep{whitney13}, 
\texttt{SPARX} (\url{https://sparx.tiara.sinica.edu.tw/})
}


\clearpage

\appendix
\section{Flow Patterns in the Hydro Models}
\label{app:flow}
\renewcommand\thefigure{\thesection.\arabic{figure}}    
\setcounter{figure}{0}    
\renewcommand\thetable{\thesection.\arabic{table}}    
\setcounter{table}{0}    

\begin{figure*}
\begin{center}
\text{\large Meridional flow fields} \par\smallskip
\includegraphics[trim=0 0 0 0, clip,width=0.8\textwidth,angle=0]{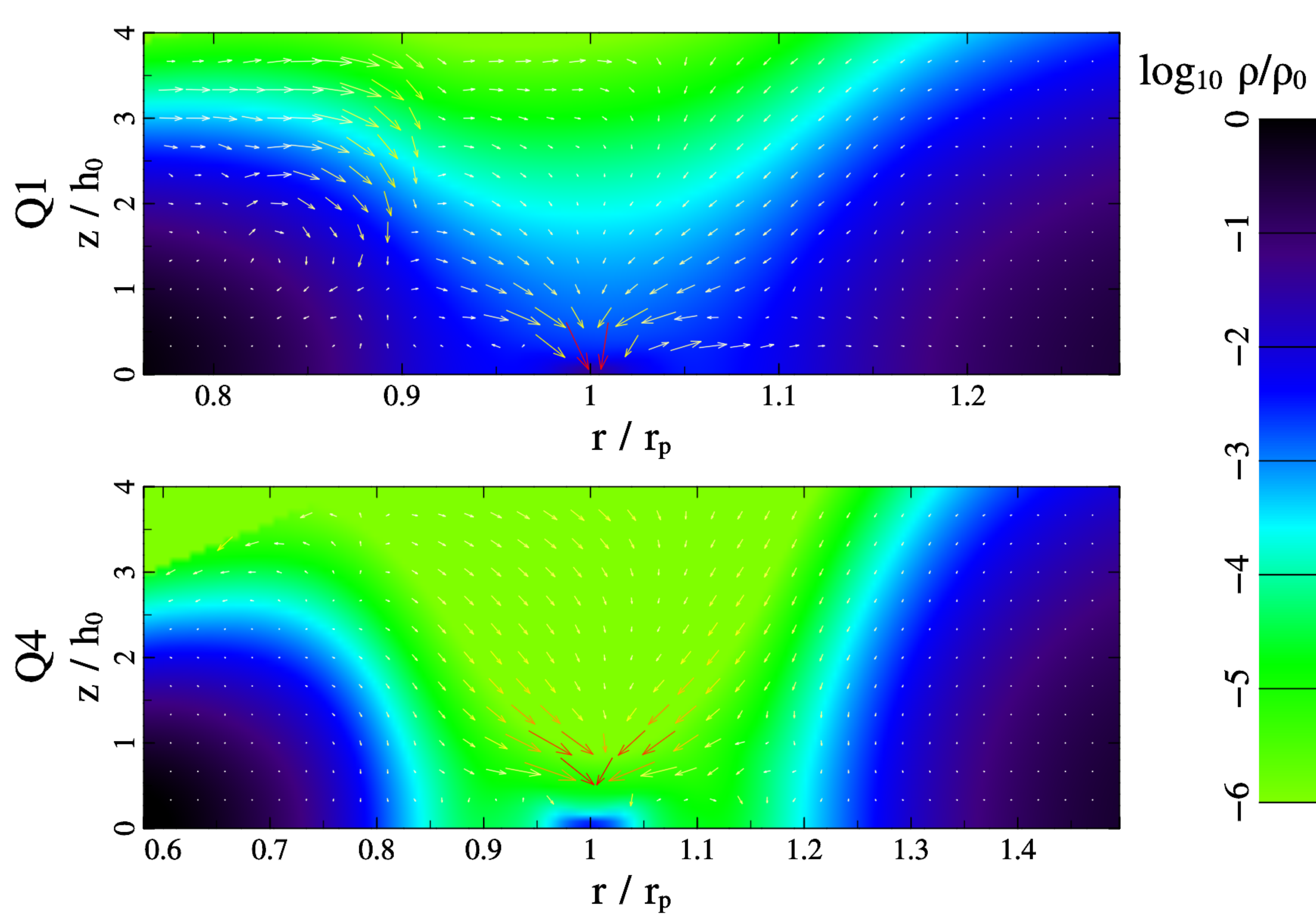}
\end{center}
\figcaption{Meridional flow fields for for Models Q1 (top) and Q4 (bottom). Arrows represent velocity vectors, and arrow lengths represent speeds. The longest arrows correspond to 0.3 $\cs$ for the top panel, and 2 $\cs$ for the bottom. Background color represents gas density, normalized to the gas density at the planet's location in an unperturbed disk. All quantities are azimuthally averaged, and time averaged from 1000 to 1010 planetary orbits. Velocity vectors within 0.5 Hill radius from the planet are omitted for clarity. 
\label{fig:v_flow_2p}}
\end{figure*}

As discussed in Section \ref{sec:hydro}, the two simulations used in this work, Models Q1 and Q4, are essentially replicas of the simulations done by \citet{fung16} with slightly modified domain size. Comparing their Figure 4 with our Figure \ref{fig:v_flow_2p}, we find the overall meridional flow pattern remains the same. Gas is repelled from the planet's gap region near the midplane, rises near the gap edge to about 2 or 3 scale heights, circulates back into the gap, falls and returns to the midplane to repeat the cycle again. 

Despite the similarities, one noticeable difference is that meridional motion in the inner and outer disks are slower in our simulations by a factor of a few. The difference is likely due to both the more extended boundaries reducing the amount of wave reflection, and the more stable advection scheme used in the updated version of \texttt{PEnGUIn}. While this is an improvement, since this work focuses on gas motion within the gap, it has little impact on our results.

\begin{figure*}
\begin{center}
\text{\large Vertical Kinetic Energy Map} \par\smallskip
\includegraphics[trim=0 0 0 0, clip,width=0.8\textwidth,angle=0]{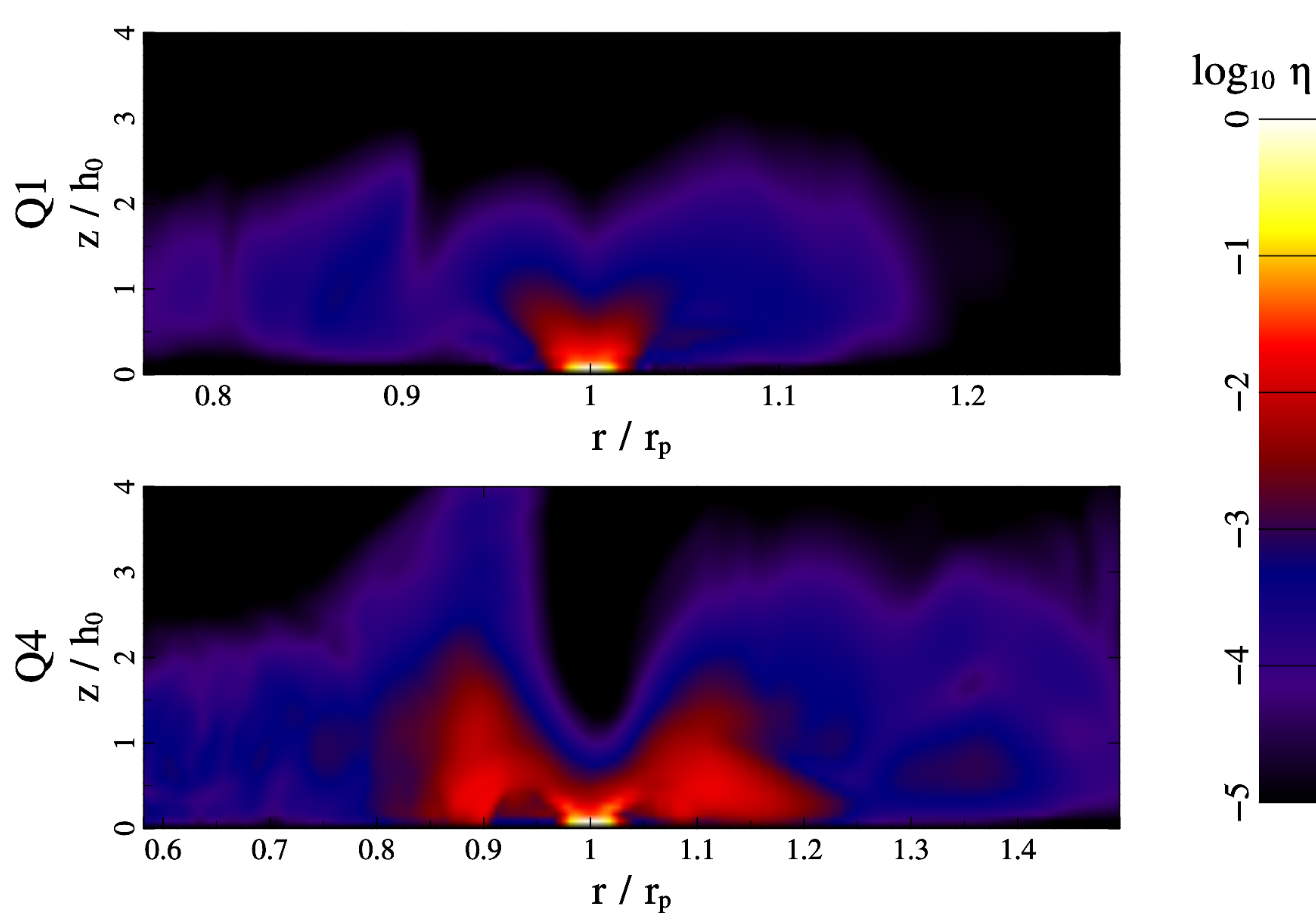}
\end{center}
\figcaption{$\eta$ (Equation \ref{eq:eta}) for Models Q1 (top) and Q4 (bottom). The quantity $\eta$ is the azimuthally averaged vertical kinetic energy density, normalized by the local disk surface density and temperature, and unlike Figure \ref{fig:v_flow_2p}, it is instantaneous at 1000 orbits without any time-averaging. 
It represents the square of the ratio of planet-induced kinematic line broadening to thermal broadening.
\label{fig:meridionalkinematicenergy}}
\end{figure*}

Figure \ref{fig:meridionalkinematicenergy} plots the following  quantity $\eta$:
\begin{equation}
\label{eq:eta}
\eta(r, z) = \frac{ \int_0^{2\pi}~\rho v_{\rm z}^2~{\rm d}\phi }{ \cs^2 \int_0^{2\pi} \int_{-\infty}^{\infty}~\rho~{\rm d}z~{\rm d}\phi } \, ,
\end{equation}
where $\cs$ is taken from Equation \ref{eq:hydro_cs}. The vertically integrated $\eta$ is analogous to moment-2 detected face-on using an dense gas tracer; therefore $\eta$ can be thought of as the vertically differentiated moment-2. Then Figure \ref{fig:meridionalkinematicenergy} essentially illustrates which portion of the gap contributes more to the kinematic fraction of moment-2. Evidently, most of the contribution comes from near the midplane. Comparing it with Figure \ref{fig:v_flow_2p}, we see that the locations with the highest speeds do not necessarily contribute most, as those regions also tend to have low densities.

\end{document}